\documentclass[12pt]{article}
\usepackage{epsf}
\usepackage{epsfig}
\usepackage{cite}

\newcommand{\bmat}{\left(\begin{array}}
\newcommand{\emat}{\end{array}\right)}

\def\yzero{\smash{\hbox{$y\kern-4pt\raise1pt\hbox{${}^\circ$}$}}}

\def\a{\alpha}

\def\d{\delta}
\def\beq{\begin{equation}}
\def\eeq{\end{equation}}
\def\beqa{\begin{eqnarray}}
\def\eeqa{\end{eqnarray}}

\def\th{\theta}
\def\vt{\vartheta}

\def\-{\hphantom{-}}

\def\s2{\frac{1}{\sqrt2}}

\def\oh{\frac{1}{2}}
\def\beq{\begin{equation}}
\def\eeq{\end{equation}}
\def\beqa{\begin{eqnarray}}
\def\eeqa{\end{eqnarray}}

\def\IF{\relax{\rm I\kern-.18em F}}
\def\II{\relax{\rm I\kern-.18em I}}
\def\IP{\relax{\rm I\kern-.18em P}}
\def\IC{\relax\hbox{\kern.25em$\inbar\kern-.3em{\rm C}$}}
\def\IR{\relax{\rm I\kern-.18em R}}

\def\cn{{\cal N}}

\def\Dsl{\,\raise.15ex\hbox{/}\mkern-13.5mu D} 
\def\IZ{Z\kern-.4em  Z}

\def\int{{\bf Z}}
\def\C{{\bf C}}
\def\ti{\times}

\def\R{{\cal R}}

\def\eps{\epsilon}

\def\sig{\sigma}

\catcode`\@=11
\newdimen\@rotdimen
\newbox\@rotbox

\def\@vspec#1{\special{ps:#1}}
\def\@rotstart#1{\@vspec{gsave currentpoint currentpoint translate
   #1 neg exch neg exch translate}}
\def\@rotfinish{\@vspec{currentpoint grestore moveto}}
\def\@rotr#1{\@rotdimen=\ht#1\advance\@rotdimen by\dp#1%
   \hbox to\@rotdimen{\hskip\ht#1\vbox to\wd#1{\@rotstart{90 rotate}%
   \box#1\vss}\hss}\@rotfinish}
\def\@rotl#1{\@rotdimen=\ht#1\advance\@rotdimen by\dp#1%
   \hbox to\@rotdimen{\vbox to\wd#1{\vskip\wd#1\@rotstart{270 rotate}%
   \box#1\vss}\hss}\@rotfinish}%
\def\@rotu#1{\@rotdimen=\ht#1\advance\@rotdimen by\dp#1%
   \hbox to\wd#1{\hskip\wd#1\vbox to\@rotdimen{\vskip\@rotdimen
   \@rotstart{-1 dup scale}\box#1\vss}\hss}\@rotfinish}%
\def\@rotf#1{\hbox to\wd#1{\hskip\wd#1\@rotstart{-1 1 scale}%
   \box#1\hss}\@rotfinish}%
\def\rotate{\@ifnextchar[{\@rotate}{\@rotate[l]}}
\def\@rotate[#1]#2{\setbox\@rotbox=\hbox{#2}\@nameuse{@rot#1}\@rotbox}

\catcode`\@=12

\topmargin
-1.5cm
\textwidth
15.5cm
\textheight
23.5cm
\oddsidemargin
0.7cm
\evensidemargin
1.2cm

\begin{document}

\makeatletter
\@addtoreset{equation}{section}
\makeatother
\renewcommand{\theequation}{\thesection.\arabic{equation}}
\pagestyle{empty}
\rightline{FTUAM-02/32; IFT-UAM/CSIC-02-55}
\rightline{\tt hep-ph/0212064}
\vspace{0.5cm}
\begin{center}
\LARGE{Towards a theory of quark masses, mixings and CP-violation 
\\[10mm]}
\large{ 
D. Cremades, L.E. Ib\'a\~nez and F. Marchesano
\\[2mm]}
\small{
 Departamento de F\'{\i}sica Te\'orica C-XI
and Instituto de F\'{\i}sica Te\'orica  C-XVI,\\[-0.3em]
Universidad Aut\'onoma de Madrid,
Cantoblanco, 28049 Madrid, Spain.
\\[9mm]}
\small{\bf Abstract} \\[7mm]
\end{center}

\begin{center}
\begin{minipage}[h]{15.0cm}
We discuss the structure of Yukawa couplings in D-brane 
models in which the SM fermion spectrum appears at the
intersections of D-branes wrapping a compact space. 
In simple toroidal realistic examples one can explicitly compute
the Yukawa couplings as a function of the geometrical data
 summing  over world-sheet instanton contributions. 
A particular simple model with a $N = 1$ SUSY spectrum and 
three quark-lepton generations is studied in some detail. 
Remarkably, one  can reproduce the observed
spectrum of quark masses and mixings for  particular choices of
the compact radii and brane locations. In order to reproduce
the smallness of  up- and down-quark masses branes should 
be located in simple geometric configurations leading to some
accidental global symmetries. 
We also find that the brane configurations
able to reproduce the observed data may be considered as
a deformation (by brane translation) of a  configuration
with   Pati-Salam gauge symmetry.
The origin of  CP-violation in this formalism is  quite
elegant. It appears as a consequence of the generic presence of 
$U(1)$ Wilson line backgrounds in the compact dimensions.  
One can reproduce the observed results for the 
CP-violation Jarlskog invariant J as long as the compact 
radii are of order of the string scale.

{\large DISCLAIMER:}
{\em This paper is going to be substantially revised.
 Althought the physics and general concepts are still valid, the Yukawa
couplings of the particular model presented in this paper have a
simpler form than discussed here, as we recently 
pointed out in hep-th/0302105. A properly revised version will be 
eventually  sent as the paper is appropriately corrected.}

\end{minipage}
\end{center}
\newpage
\setcounter{page}{1}
\pagestyle{plain}
\renewcommand{\thefootnote}{\arabic{footnote}}
\setcounter{footnote}{0}


\section{Introduction}
One of the most outstanding puzzles of the standard model (SM)
is the structure of fermion masses and mixing angles. 
The masses of quark and leptons are clearly not random,
showing a hierarchical structure (see table \ref{spec})
with masses differing by  several  orders of magnitude.
\begin{table}[htb] \footnotesize
\renewcommand{\arraystretch}{1.25}
\begin{center}
\begin{tabular}{|c|c|c|c|}
\hline\hline  U-quarks   &   u  &  c  &   t  \\
\hline   &
 0.9-2.9\ MeV  &  530-680\  MeV &  168-180\ GeV \\
\hline\hline
  D-quarks &  d  &  s   &   b \\
\hline
  &   1.8-5.3\ MeV  &  35-100\ MeV &  2.8-3.0 \ GeV \\
\hline\hline
  Leptons &  e  &  $\mu $  &  $\tau $  \\
\hline
  &  0.51 \ MeV &  105.6 \ MeV  &  1.777 \ GeV \\
\hline\hline
\end{tabular}
\end{center}
 \caption{\small Masses of quarks and leptons at the $M_Z$ scale
taken from the first paper in ref. \cite{fritzsch}.}
\label{spec}
\end{table}
Something similar happens with the electroweak CKM 
mixing matrix which numerically 
is given (for the moduli of the entries) by \cite{pdg}
\beq
\begin{array}{cc}
&
\\
\left| V_{CKM} \right|\ =\ 
& \left(
\begin{array}{ccc}
 0.9741-0.9756  &  0.219-0.226  &  0.0025-0.0048 \\
0.219-0.226  & 0.9732-0.9748  &   0.038-0.044 \\
0.004-0.014 & 0.037-0.044 &  0.9990-0.9993
\end{array}
\right)  \ . 
\end{array}
\label{CKM}
\eeq
at the 90\% confidence level.
This is close to a unit matrix with small off-diagonal mixing
except for the Cabibbo (12) entry which is somewhat larger.
Again there is a hierarchical structure with the third generation
mixing mostly with the second generation rather than the first.
The violation of CP is given in the ``standard'' parametrization 
\cite{pdg} by
the phase $\delta _{13} = 59^{ 0} \pm 13^{ 0}$, which may be considered 
rather large. A convention independent measure of CP-violation
\cite{jarlskog} 
is given by the Jarlskog invariant J which experimentally
is \cite{pdg} $J = (3.0\pm 0.3)\times 10^{-5}$.

The understanding of the structure of fermion masses and
mixings has been the subject of an enormous amount of 
effort. A  phenomenological attitude widely 
followed is to consider fermion mass ``textures''  or
definite ansatze for the form of the mass matrices
(see, e.g.,\cite{fritzsch} and references therein). In the
presence of ``texture zeros'' one can obtain relationships
between the mixing angles and quark masses like the
approximate numerical equality 
\cite{chamba}
$sin \ \theta_{12} = \sqrt{m_d/m_s}$.
Attempts to understand the presence of such ``textures''
have been made in different flavour models, one of the 
simplest being based in Abelian flavour symmetries
\cite{Abelian,ir}.
Grand unified theories combined with  textures have also
been systematically explored in order to understand the 
fermion spectrum (see, e.g., \cite{raby} and references therein).
In spite of the fact that these and other approaches 
can give a semiquantitative understanding of the 
fermion spectrum, we are still lacking 
a full explanation of the masses and mixings in terms of
a more fundamental theory.
An obvious candidate for such a more fundamental theory
is string theory. In fact 
the structure of fermion masses has   been studied in
a number of  semirealistic heterotic string models.
  Although
one can reproduce interesting features of the observed spectrum,
one  limitation of this approach is that  
usually the obtained models have 
(at least before field theory flat directions are taken) 
additional fermions, gauge and
Higgs bosons which make the analysis complicated.

In spite of that,
string theory has a priori a theoretical advantage for
addressing this puzzle: the Yukawa couplings may be 
computable as functions of the geometrical moduli
in particular models. That is the case, for example 
of Abelian $\int_N$ orbifold compactifications in which
one can use conformal field theory (CFT) techniques 
to compute them \cite{hv} . 
 In fact, string theory contains built-in 
a possible mechanism in order to obtain hierarchical
Yukawa couplings \cite{orbihier}.
 In such orbifold models some matter 
fields are localized in extra dimensions in the different 
fixed points of the orbifold. Thus one can imagine an
scenario in which the Higgs field lives in one of the fixed 
points and the different right- and left-handed fermions 
live in different distant fixed points. Yukawa couplings 
between  quarks which are distant from the Higgs are 
mediated by world-sheet instanton effects and are 
hence exponentially suppressed by the (distance)$^2$ 
between the different fixed points \cite{orbihier} .
 The corresponding
Yukawa couplings may be explicitly computed in terms
of the geometry of the given orbifold. 
Although from the theoretical point of view this is fine,
phenomenologically the problem arises because in
such $\int_N$ heterotic orbifolds searches for
models with just three generations and a minimal 
Higgs sector have been unsuccessful.

With the advent of D-branes in Type II and type I string theory,
the phenomenological possibilities of string theory have 
widened in several respects (see ref. \cite{aiqu,bw,bgkl,
afiru,afiru2,bkl,imr,bklo,csu,bailin,cim12,hon,
koko,ellis,cim3,bbkl,angel,bgo} for
the construction of
explicit D-brane semirealistic string configurations).  
 In particular, there is  now the possibility of following a ``bottom-up''
approach \cite{aiqu,angel} 
to the construction of semirealistic three-generation models,
i.e., one can consider local D-brane configurations giving rise 
to the gauge group of the SM and three quark-lepton generations
and then embed such a local configuration in some general 
compactification. Many properties  of the models depend only on the local
brane configuration, rather than on the details of the
particular compact (e.g., Calabi-Yau) space.

In the present paper we address the computation of Yukawa couplings in 
D-brane models. We concentrate on
a recently constructed class of D-brane models 
\cite{bgkl,afiru,afiru2,bkl,imr,bklo,csu,bailin,cim12,hon,
koko,ellis,cim3,bbkl,angel,bgo}
in which the SM chiral spectrum is obtained at intersecting branes
(for short reviews see, e.g., \cite{dubna,hamburg}).
Nevertheless, we believe the approach 
is much more general since Yukawa couplings in other type of
D-brane models (like those based on D-branes at orbifold 
singularities or those involving D-branes with magnetic fluxes
\cite{aads}) are related by T-duality and/or mirror 
symmetry with intersecting brane configurations \cite{torons,angel}.

\begin{figure}[ht]
\centering
\epsfxsize=2.9in
\hspace*{0in}\vspace*{.2in}
\epsffile{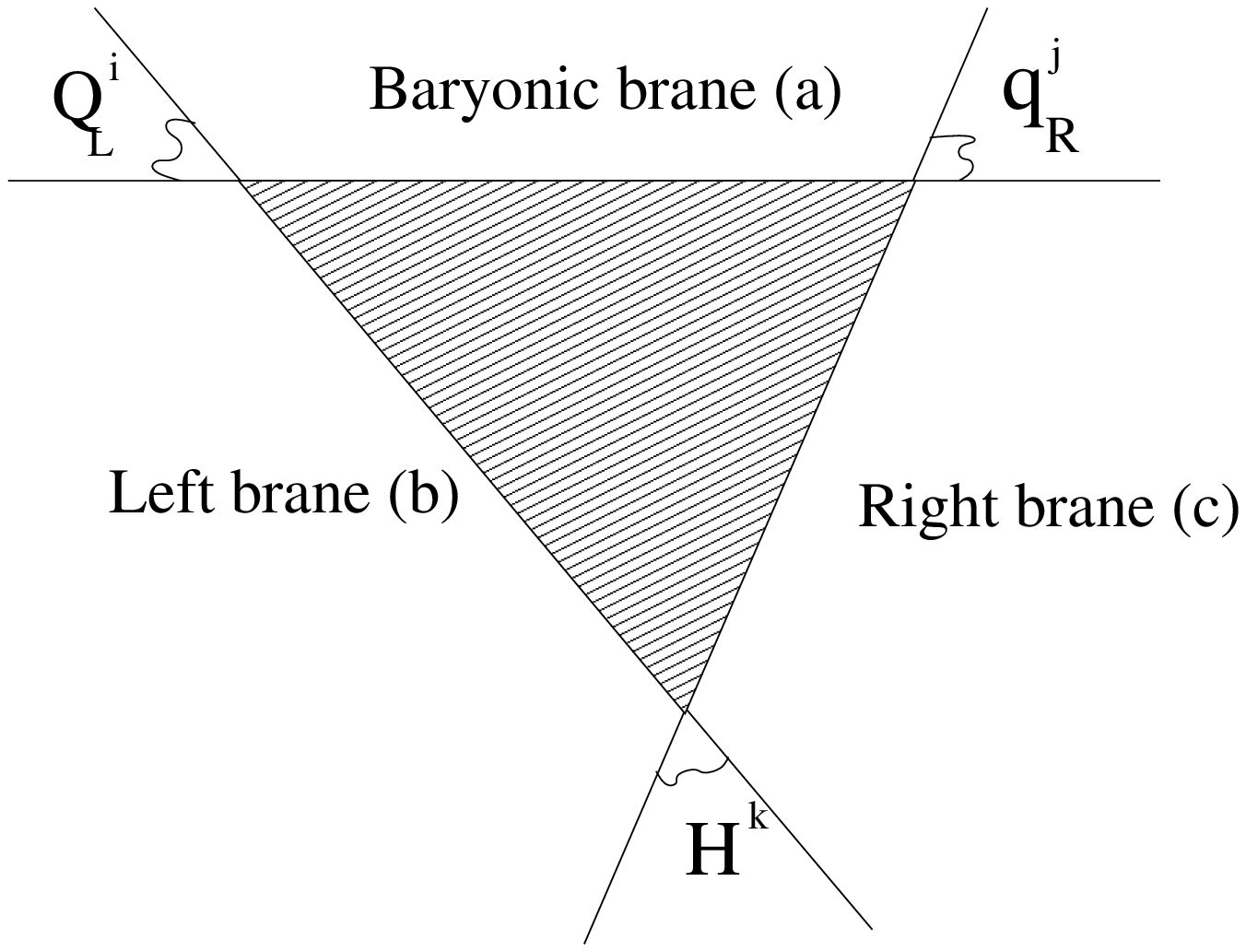}
\caption{\small{Yukawa coupling between two quarks of opposite chirality
and a Higgs boson.}}
\label{tri1}
\end{figure}

One of the  advantages of the present  approach is that, 
unlike what happened in heterotic constructions, 
it is relatively easy to
find D-brane string configurations yielding 
(at least for the chiral spectrum) just the particle content
of the SM (or the MSSM) {\it already at the string level}, i.e.,
without any further effective field theory elaboration \cite{imr,cim12}. 
The simplest such models involve orientifold compactifications of
Type IIA string theory on a 6-torus $T^2\times T^2\times T^2$.
The Yukawa couplings in these schemes correspond to string correlators 
involving the Higgs scalar, one right-handed and one left-handed 
fermion \cite{afiru2}. The worldsheets corresponding to those correlators 
have then a triangular shape with the fields at the intersections 
and the sides embedded on the three different intersecting stacks of
branes
(see figure \ref{tri1}). The leading contribution to the 
Yukawa couplings is  then given by worldsheet instanton 
contributions proportional to $exp(-S_{cl})= exp(-A_{abc})$, where
$A_{abc}$ is the area of the triangles \cite{afiru2}. 
In this paper we will focus mainly on the phenomenological
applications of such computations and leave the derivation of 
general expressions for the Yukawa couplings for 
arbitrary toroidal Dp-brane configurations for a separate paper
\cite{cim5}.

In order to show the possibilities
of the  approach, we compute the Yukawa couplings for a
new simple  model yielding
the chiral spectrum of the MSSM at the intersection
of D6-branes wrapping a 6-torus
\footnote{A similar analysis may be attempted with other
intersecting brane models like those in refs.
\cite{imr,cim12}. The present
model has the advantage of having a minimal Higgs sector,
 $\cn = 1$ SUSY at all intersections
and quite a simple geometrical configuration.
This allows for rather simple expressions for the Yukawa couplings.}.
 We provide explicit simple formulae 
for the relevant Yukawa couplings in terms of world-sheet 
instanton sums over the cycles of the 6-torus. In some particular
geometrical configurations the Yukawa couplings may be written as
products of Jacobi $\theta$-functions with characteristics.
One of the nicest features we find is 
 a natural origin for the complex phases 
necessary for CP-violation. In the present approach they correspond to
the generic presence of non-trivial Wilson line backgrounds 
associated to the Abelian gauge symmetries of the theory.

We have performed a search for particular D-brane geometries
in this model capable of describing the observed 
quark masses, mixings and CP-violation parameters.
We have found that for certain torus radii and D6-brane
locations one can reproduce the observed quark masses
and mixing angles. On the other hand we also find  the
interesting result that, in order to reproduce the 
observed size for CP-violation, the compact radii 
of two of the tori have  to be of order the string scale,
otherwise the phases get exponentially suppressed 
as the radii increase. 
The D6-brane locations 
reproducing the observed data are close to 
positions in the tori with certain symmetry properties.
At those symmetric positions pairs of triangles 
corresponding to different quark generations become equal, 
signalling some accidental global symmetries.

Finally we find that the observed  spectrum of fermion masses 
and mixings is consistent with the existence of an underlying 
(broken)  Pati-Salam symmetry \cite{ps}. More precisely, the brane configurations
able to reproduce the observed data may be considered as 
a deformation (by brane translation) of a  configuration
leading to a Pati-Salam symmetry.

The structure of the rest of the paper is as follows. In the next chapter
we give a brief introduction to intersecting brane models and 
present the specific model to be analyzed numerically afterwards.
In chapter three we present the form of the expressions for the
Yukawa couplings  and discuss the origin of
symmetries and CP-violation. In chapter four we perform 
a numerical analysis and show how all quark masses, mixings
and CP-violation parameters may be reproduced with the given 
Yukawa coupling formulae. We also extend the discussion 
to charged lepton and (Dirac) neutrino masses, mixings and
CP-violation.  We leave chapter five for some final comments  and 
conclusions.

\section{Intersecting brane standard models}

\begin{figure}[ht]
\centering
\epsfxsize=5.5in
\hspace*{0in}\vspace*{.2in}
\epsffile{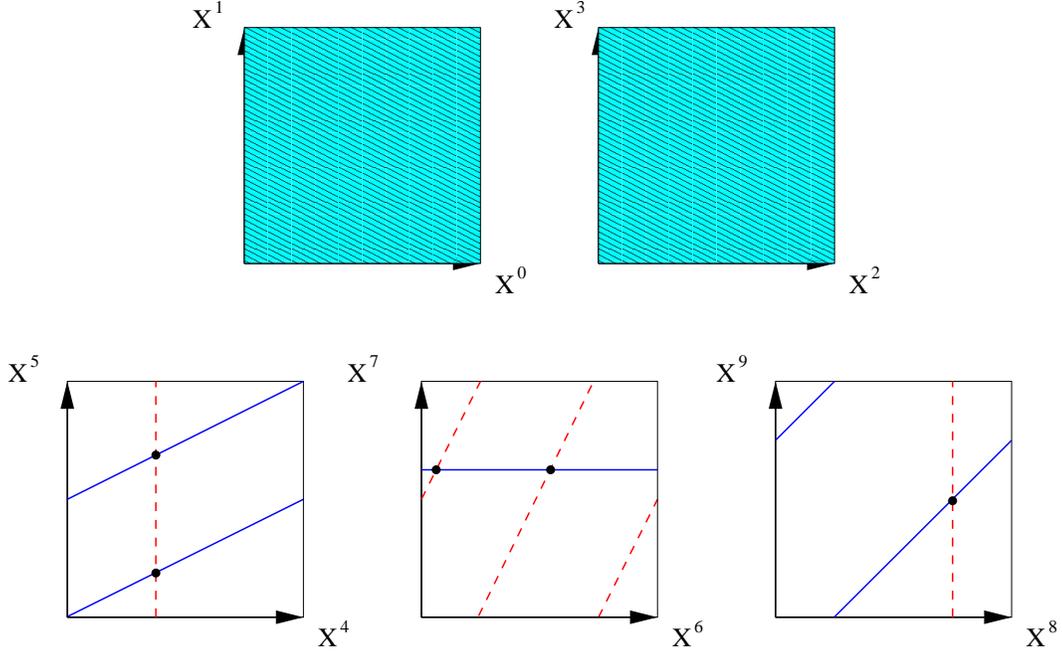}
\caption{\small{Intersecting brane world setup. Consider two D6-branes  
filling four non-compact dimensions, to be identified with $M_4$, 
and wrapping three 1-cycles of $T^2 \ti T^2 \ti T^2$.  
In this example the wrapping numbers are $(2,1)(1,0)(1,1)$ 
(solid line) and $(0,1)(1,2)(0,1)$ (dashed line).
The total intersection number is $2 \times 2 \times 1 = 4$.
}}
\label{world}
\end{figure}

A particular class of models with intersecting D-branes 
has received recently considerable attention,
\cite{bgkl,afiru,bkl,imr}. These are 
models in which one has four stacks of intersecting D-branes 
in an orientifold of Type IIA string theory. The 
four stacks come under the names of  {\it 
baryonic, left, right} and {\it leptonic }  and each stack
gives rise to the gauge groups $SU(3)$, $SU(2)_L$, $U(1)_R$ 
and $U(1)_{lepton}$ respectively (with possibly additional 
$U(1)$'s). Open strings at the intersections give rise
to chiral fermions with the quantum numbers of
SM quarks and leptons. In one of the simplest class of
models the Type IIA string is compactified on a
6-torus $T^2\times T^2\times T^2$ and D6-branes wrap
some cycles of the tori. 
Specifically, the 7-dimensional worldvolume
of D6-branes contains Minkowski space and the three 
remaining dimensions of the D6-branes 
wrap one-cycles at each of the three 2-tori of the
$T^6$ (see figure \ref{world}). Let us  
denote by $(n_a^i, m_a^i)$, $i$ = 1,2,3
the wrapping numbers of each brane $D6_a$, $n_a^i$ ($m_a^i$) being the
number
of times the brane stack $a$  
is wrapping around the  $x$($y$)-coordinate of the 
$i^{th}$ torus. One can check that the number of times two branes
$D6_a$  and $D6_b$ intersect in $T^6$ is given by the intersection number
\cite{bgkl}:
\beq
I_{ab}\ =\
(n_a^1m_b^1-m_a^1n_b^1)(n_a^2m_b^2-m_a^2n_b^2)(n_a^3m_b^3-m_a^3n_b^3)
\label{internumber}
\eeq
%
Open strings stretching around the intersections give rise to
chiral fermions in the bifundamental representation $(N_a,{\overline
N}_b)$
under the gauge group of the two branes $U(N_a)\times U(N_b)$.
Thus, these configurations  yield $I_{ab}$ copies of the same
bifundamental representation, providing a natural source for the observed
generation replication.
In fact, because of technical reasons\footnote{In the example below 
the orientifold operation is required, e.g., in order to get a non-Abelian 
$SU(2)$ group for left-handed weak interactions.} one performs 
an ``orientifold'' twist
\cite{orientifold} of this theory by the product
$\Omega \times \R$, where $\Omega$ is the worldsheet parity operator
and $\R$ is the reflection  with respect to the 
horizontal  axis of the three tori. 
Now, for consistency of the construction, one has to include 
in the D6-brane configuration {\it mirror branes} $D6_{a*}$ for each
original $D6_a$-brane, wrapping geometrical loci that
are mirror with respect to the reflection operation $\R$.
In particular, this implies that their wrapping numbers will be 
$(n_a^i, - m_a^i)$, rather than $(n_a^i, m_a^i)$.
Being part of the same dynamical object, both $D6_{a}$ and 
$D6_{a*}$ branes correspond to the same gauge group, generically
$U(N_a)$. The chiral fermions arising from the intersection $ab$*,
though, will transform in the bifundamental representation $(N_a,N_b)$,
rather than $(N_a,{\overline N}_b)$. Finally, when branes $a$ and $a$* 
coincide, the gauge group may be enhanced from $U(N_a)$ to $SO(2N_a)$
or $USp(2N_a)$.

Let us now show a particular D-brane configuration giving rise to
three quark-lepton generations. This model has not been studied 
previously and we chose  it here because of its simplicity
and because it naturally contains a minimal Higgs sector
which makes very simple the study of Yukawa couplings. 
Consider the stacks of D6-branes with the wrapping numbers of
table \ref{MSSMwrappings}.
\begin{table}[htb] \footnotesize
\renewcommand{\arraystretch}{1.5}
\begin{center}
\begin{tabular}{|c||c|c|c|}
\hline
 $N_i$    &  $(n_i^1,m_i^1)$  &  $(n_i^2,m_i^2)$   & $(n_i^3,m_i^3)$ \\
\hline\hline $N_a=3$ & $(1,0)$  &  $(1/\rho, 3\rho)$ &
 $(1/\rho, -3\rho )$  \\
\hline $N_b=1$ &   $(0, 1)$    &  $ (1,0)$  &
$(0,-1)$   \\
\hline $N_c=1$ & $(0,1)$  &
 $(0,-1)$  & $(1,0)$  \\
\hline $N_d=1$ &   $(1,0)$    &  $(1/\rho, 3\rho)$ &
 $(1/\rho, -3\rho )$ \\
\hline \end{tabular}
\end{center} \caption{\small D6-brane wrapping numbers giving rise to
the chiral spectrum of the MSSM. The discrete parameter 
$\rho = 1, 1/3$ describes two different sets of 
wrapping numbers yielding the same intersection numbers (\ref{intersecMSSM}).
The addition of the mirror branes is understood. 
\label{MSSMwrappings} }
\end{table}
Generically, the gauge group of this configuration is
$U(3)\times U(1)^3$. However, one can check that the
symmetry is enhanced to $U(3)_a\times SU(2)_b\times U(1)_c\times U(1)_d$
if the brane $b$ is located on top of its orientifold mirror
$b$*. Computing the intersection numbers as above one gets
the result

\beq 
\begin{array}{lcl} 
I_{ab}\  =  \ 3, & &  I_{ab*}\ =\ 3,  \\
I_{ac}\  =  \ -3, & &  I_{ac*}\ =\ -3, \\
I_{db}\  =  \ 3, & & I_{db*}\ =\ 3,  \\
I_{dc}\  =  \ -3, & & I_{dc*}\ =\ 3,  \\
I_{bc}\  =  \ -1, & & I_{bc*}\ =\ 1,
\end{array}
\label{intersecMSSM}
\eeq
which corresponds to the
 chiral fermion spectrum of the SM (plus right-handed neutrinos,
see table \ref{mssm}).
In addition there is a minimal set of Higgs multiplets if
one locates the brane $b$ on top of the brane $c$ along the first
torus (otherwise the branes do not intersect 
and the state is massive). In other words, there is a
minimal Higgs sector with a $\mu$-parameter 
whose real part is given by the distance between branes
$b$ and $c$ along the first torus 
(see figure \ref{guay} below)
and its imaginary part is proportional
to the relative phase of their Wilson lines
\footnote{By a Wilson line we mean a constant background gauge potential
$A_\sig (x_\sig^\a)$ living on the worldvolume of the D6-brane $\sig$, 
and that depends on the three internal coordinates $x_\sig^\a$ that wrap 
$T^6$. Since each of these internal coordinates have the 
topology of a non-contractible circle, $A_\sig$ cannot be gauged away. 
We will take $A_\sig(x_\sig^\a)$ to take values in a $U(1)$ 
subgroup of the full gauge group for each stack $\sig$. In this way, when
an open string endpoint attached to the brane $\sig$ makes a closed loop 
in the $\a^{th}$ torus, its associated wavefunction picks up a phase 
$exp \ (i 2 \pi \theta_\sig^{(\a)})$.\label{Wilson}}
 in the same torus.
\begin{table}[htb] \footnotesize
\renewcommand{\arraystretch}{1.25}
\begin{center}
\begin{tabular}{|c|c|c|c|c|c|c|}
\hline Intersection &
 Matter fields  &   &  $Q_a$   & $Q_c $ & $Q_d$  & Y \\
\hline\hline (ab),(ab*) & $Q_L$ &  $3(3,2)$ & 1   & 0 & 0 & 1/6 \\
\hline (ac) & $U_R$   &  $3( {\bar 3},1)$ & -1  &  1  & 0 & -2/3 \\
\hline (ac*) & $D_R$   &  $3( {\bar 3},1)$ &  -1  & -1    & 0 & 1/3
\\
\hline (db),(db*) & $L$    &  $3(1,2)$ &  0    & 0  & 1 & -1/2  \\
\hline (dc) & $N_R$   &  $3(1,1)$ &  0  & 1  &  -1  &  0  \\
\hline (dc*) & $E_R$   &  $3(1,1)$ &  0  & -1  & -1    & 1 \\
\hline (cb),(cb*) & $H_{u,d}$   &  $(1,2)$ &  0 &  $\pm 1$ & 0  &
$\mp 1/2$  \\
\hline \end{tabular}
\end{center} \caption{\small Standard model spectrum and $U(1)$ charges.
The
hypercharge
generator is defined as $Q_Y = \frac 16 Q_a - \frac 12 Q_c - \frac 12
Q_d$.}
\label{mssm}
\end{table}

If the ratios of radii in the second and third tori are equal 
($R_2^{(2)}/R_1^{(2)}=R_2^{(3)}/R_1^{(3)}=\chi$)
one can check that  the same $\cn = 1$ SUSY  is preserved at all
intersections \cite{csu,cim12}. So this configuration is (locally) 
$\cn = 1$
supersymmetric, and  the massless chiral spectrum is that
of the MSSM with a minimal Higgs set.
 In this model there are three $U(1)$'s and only
one of them $(3B+L)$ is anomalous. As usual in this class
of models (see, e.g., ref.\cite{imr,uunos}) the anomalous 
$U(1)$ gets massive
by combining with one RR-field. There are two massless $U(1)$'s
corresponding to $(B-L)$ and the 3$^{rd}$ component of right-handed
weak isospin ($U(1)_c$). So the actual low-energy gauge group
is $SU(3)\times SU(2)\times U(1)_{B-L}\times U(1)_c$.
The model may be further broken to the SM e.g., by
giving a vev to the right-handed sneutrino. This may be triggered by the 
presence of a FI-term for the anomalous $U(1)$, along the lines explained 
in ref.\cite{cim12}. We will not, however, discuss this issue any longer 
here, since our main interest will be on the 
computation of the standard Yukawa couplings of Higgs doublets
to quarks and leptons.

%
\begin{figure}[ht]
\centering
\epsfxsize=6in
\epsffile{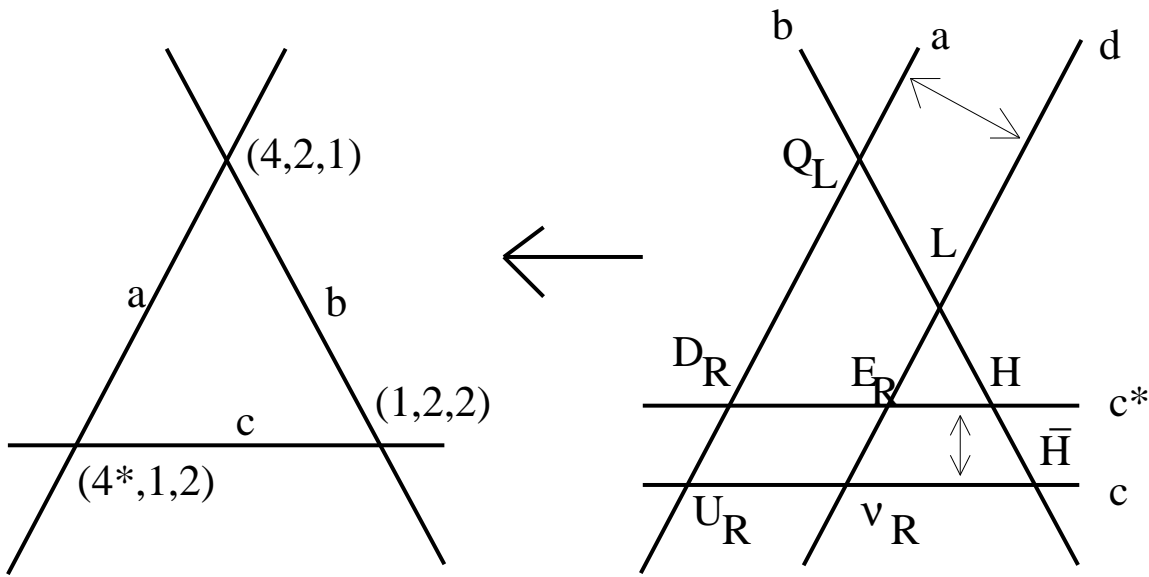}
\caption{\small{ Scheme of the 
model in the text. Moving brane $d$ on top 
of brane $a$ one gets an enhanced $SU(4)$ Pati-Salam symmetry.
If in addition brane $c$ is located on top of its mirror
$c$* there is an enhanced $SU(2)_R$ symmetry.}}
\label{guayps}
\end{figure}

Note that the gauge interactions of this brane configuration
may be enhanced to a full Pati-Salam $SU(4)\times SU(2)_L\times SU(2)_R$.
Indeed, branes $a$ and $d$ are parallel (see table 
(\ref{MSSMwrappings}) )and if we put 
the leptonic brane on top of the baryonic branes the gauge
group is enhanced $SU(3) \times U(1)_{B-L} \rightarrow SU(4)$.
Moreover if the brane $c$ is put on top of its mirror $c$*,
there is the further enhancement $U(1)_c\rightarrow SU(2)_R$.
Thus, the model we are studying may be considered
as coming from an adjoint breaking of a Pati-Salam 
model \footnote{ In these brane configurations adjoint Higgsing 
corresponds to parallel brane separation and/or 
non-trivial Wilson lines along a non-Abelian subgroup.}

%
\begin{figure}[ht]
\centering
\epsfxsize=6.5in
\epsffile{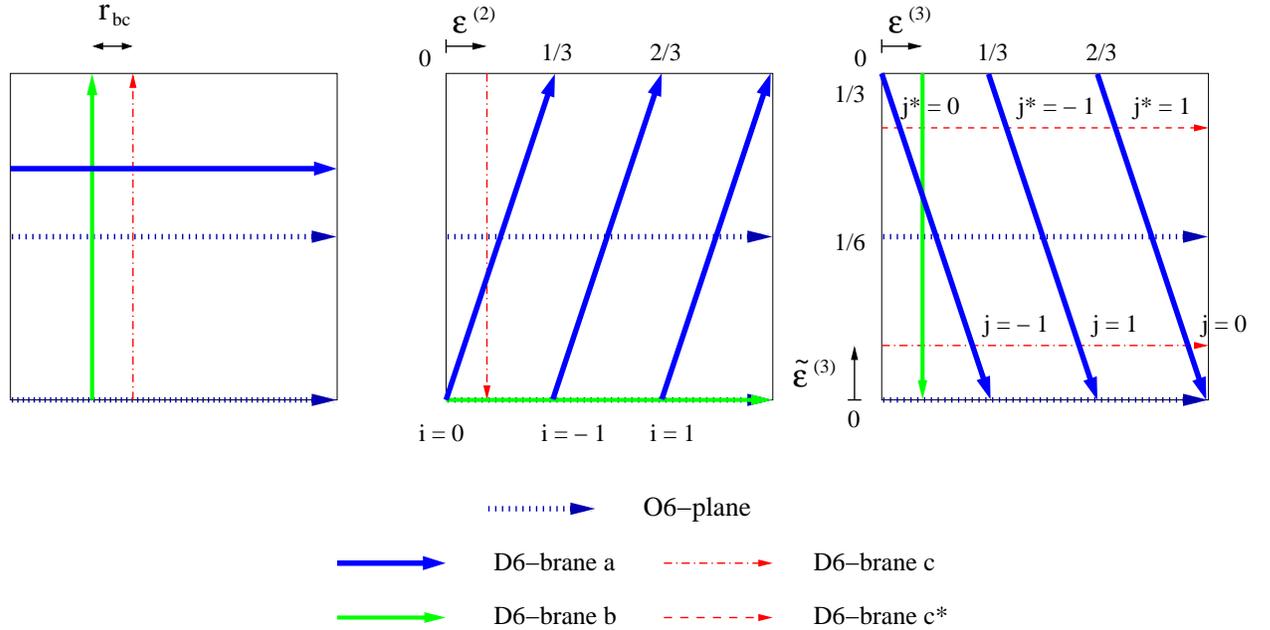}
\caption{\small{ Brane configuration corresponding to the
MSSM-like model described in the text, for the specific value $\rho = 1$. 
For simplicity, we have not depicted the leptonic brane nor the mirror
$a$* brane.}}
\label{guay}
\end{figure}

Before entering into the details of the Yukawa couplings, let us 
give a geometric description of the D-brane model under configuration,
with the wrapping numbers given in table \ref{MSSMwrappings}.
Here, we will concentrate on the choice of parameter $\rho = 1$,
but a equivalent description can be given for the model with
$\rho = 1/3$.
The distribution of the $a$, $b$ and $c$ branes is displayed in figure
\ref{guay}. We have not included the leptonic brane (which is anyway
parallel to the baryonic one) for clarity.
The figure shows the unit cells of the three 2-tori which have been
chosen here square for simplicity. 
With an horizontal discontinuous arrow is denoted the 
``orientifold hyperplane'', i.e., the region of space 
which is left fixed under the orientifold reflection $\Omega \times \R$
with respect to the horizontal axis 
(note that the horizontal axis itself is also an orientifold hyperplane).
Note that the brane $b$, associated to electroweak
$SU(2)_L$, is taken on top of an orientifold plane 
so that branes $b$ and $b$* coincide geometrically and the 
gauge group is enhanced from $U(1)$ to 
$USp(2) \simeq  SU(2)$. The other straight lines 
represent the different wrapping branes. Thus, e.g., the 
thick line represents the baryonic stack which wraps once
around the horizontal cycle in the first torus. 
In the second torus it wraps three times around the vertical cycle
and once on the horizontal axis. Without loss of generality we have 
chosen the cycle to pass through the origin. Finally, this baryonic stack
wraps three times around the vertical axis (with opposite orientation)
and once around the horizontal one in the third torus.
 Note that this baryonic cycle  
intersects three times the $b$, $c$ and $c$* branes, giving rise at those 
intersections to the three generations of quarks.
The left-handed quarks ($ab$ intersections) are
labelled by the index $i = 0, -1, 1$ and the origin of triplication
lies in the second torus. The right-handed U-quarks
(coming from $ac$ intersections) are labelled by $ j = -1, 1, 0$
and their triplication takes place in the third torus.
The same phenomenon occurs for the right-handed D-quarks
(now located at $ac$* intersections) 
which are labelled by $j$* $= -1, 1, 0$.
This distribution of intersections turns out to be 
important for the properties of the corresponding Yukawa couplings.  

Note that the branes $b$ and $c$ are parallel on the first torus
(although they intersect at the other two tori). If we put them
on top of each other in the first torus, massless chiral
multiplets with the quantum number of Higgs fields of the MSSM
appear. As we separate those two branes in the first 
torus, open strings have to stretch and the Higgs doublets 
get a mass $r_{bc}/(2\pi \alpha ')$, where $r_{bc}$ 
is the distance between those branes in the first torus.
Alternatively, if we put branes $b$ and $c$ on top of each other
but we turn on Wilson lines on the circle that they wrap on the 
first torus, strings between $b$ and $c$ will pick up a phase
$exp (2\pi i \varphi_{bc})$, with
$\varphi_{bc} = \theta_b^{(1)} - \theta_c^{(1)}$ , when
going around such circle, so that their quantized momenta
will be shifted and this will induce a mass 
$2 \pi\varphi_{bc}/ R_2^{(1)}$.
Thus, both the distance $r_{bc}$ and the relative Wilson line
$\varphi_{bc}$ are related to the $\mu$-parameter 
in this configuration. Actually, they are proportional
to the real and imaginary part of $\mu$, respectively. 

Notice finally that for ${\tilde \epsilon}^{(3)} = 0, 1/6$, the 
brane $c$ lies on top of its mirror $c$* and there is 
an enhancement $U(1)_c\rightarrow SU(2)_R$. Thus we 
recover a left-right symmetric model 
$SU(3)\times SU(2)_L\times SU(2)_R \times U(1)_{B-L}$.
If in addition the leptonic brane sits on top
of the baryonic stack (i.e.,  
$\epsilon_l^{(2)} = \epsilon^{(2)}$ and
$\epsilon_l^{(3)} = \epsilon^{(3)}$) the gauge symmetry
is further enhanced to a full Pati-Salam symmetry
$SU(4)\times SU(2)_L\times SU(2)_R$.
It is interesting to notice that in such a model with the minimal 
Higgs sector quark and lepton masses are unified and, due to
the left-right symmetry the U- and D-quark mass matrices 
are proportional an hence there is no mixing.
Separating the branes as in the model we are considering 
breaks the symmetry down to $SU(3)\times SU(2)_L\times U(1)^2$.
Thus the distance of the separated branes controls how far
away is the model from a PS symmetry.

\section{ The Yukawa couplings and their symmetries}

\subsection{The Yukawa couplings}

 In intersecting brane world configurations, chiral matter arises
from brane intersections, being localised in a internal compact manifold.
As discussed in \cite{afiru2}, a Yukawa coupling arises from  open strings
stretching a worldsheet with triangle shape in which 
chiral fields lie at the three vertices (see also \cite{mirjam}). 
The corresponding semiclassical amplitude  to the Effective
Lagrangian has the form
\beq
h_{ijk} \sim {\rm exp} \left( - {A_{ijk} \over 2\pi \a^\prime} \right),
\label{yuki}
\eeq
where $A_{ijk}$ is the  area connecting the $i,j$ and $k$
intersections.
In realistic models we will consider, for instance, three different stacks
of branes $a$, $b$ and $c$, such that all of them intersect each other, 
possibly several times. Let us suppose that $ab$ intersections yield
left-handed quarks, $ac$ intersections yield right-handed quarks and
finally
$bc$ intersections give us Higgs bosons. Then, (\ref{yuki}) would give us
a contribution to the
corresponding quark-Yukawa for each triplet of intersections of each type,
as depicted in figure \ref{tri1}.

\begin{figure}[ht]
\centering
\epsfxsize=6in
\epsffile{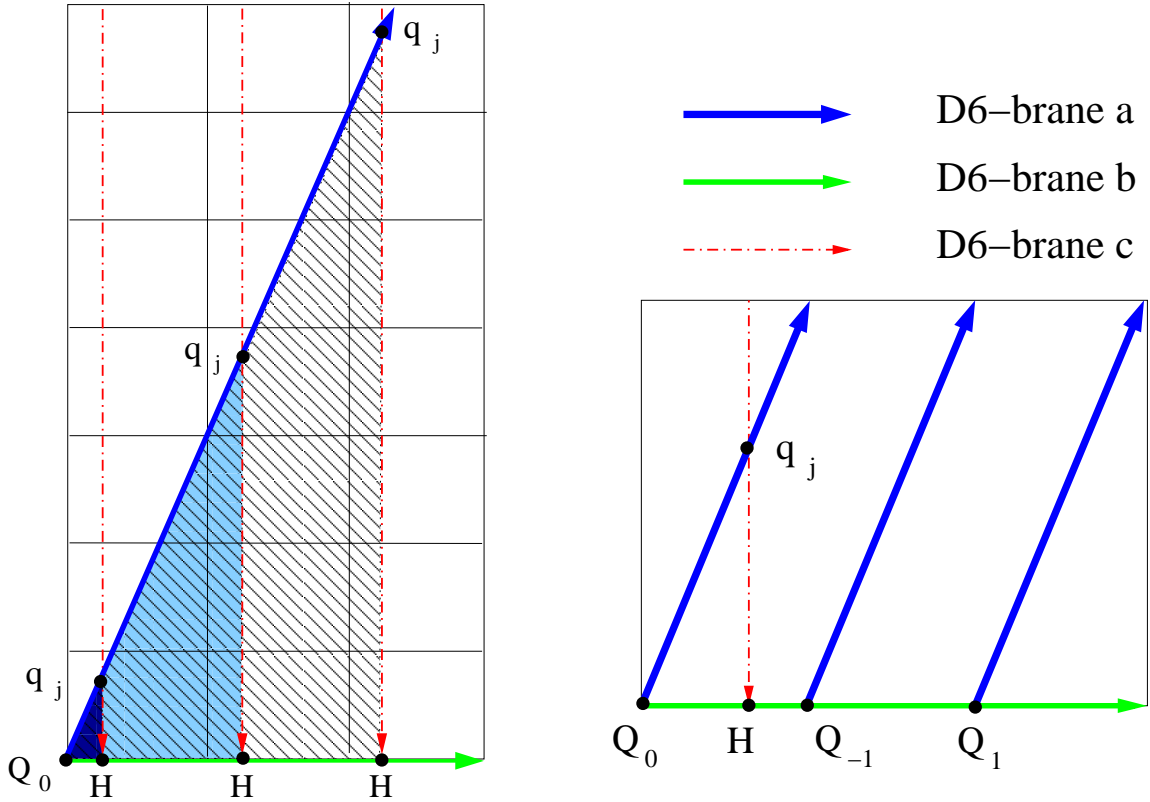}
\caption{\small{ 
Different triangular instantons contributing to a Yukawa coupling. 
In a compact space like a $T^6$, three intersection
points are connected by an infinite number of triangles, each of them giving
rise to a contribution to the {\em same} Yukawa coupling. We have 
illustrated this effect in the simple case of a two-torus (namely, the second
two-torus of figure \ref{guay}). In the figure at the right we have 
depicted a fundamental region of the torus, where in the left we have
patched several copies together. There we have drawn three triangles
of increasing area connecting the same intersection points, and thus
contributing to the same Yukawa $Q_0 H q_j$. }}
\label{solitons}
\end{figure}

In toroidal, orbifold and orientifold compactifications, such minimal
surfaces are given by triangles whose vertices are at the intersections
and whose sides lie on the worldvolume of the branes. The value of such areas
is then, in principle, computable (at least at a classical level)
in terms of the geometrical data of the brane configuration.
In fact, due to the compact toroidal geometry, for a given Yukawa
coupling,
there is more than one triangle contributing to the amplitude since there
are also triangles wrapping a number of times the tori 
(see fig. \ref{solitons}). 
Thus, actually the Yukawa coupling comes as a worldsheet instanton sum:
\beq
h_{ijk}  \sim   \sum_{l_1,l_2,l_3} e^{- \frac{ A_{ijk} \left(l_1,l_2,l_3
\right) }{2 \pi \a^\prime}
}
 e^{ -2\pi i \phi \left(l_1,l_2,l_3 \right)} 
\label{yukiguay}
\eeq

For a given Yukawa coupling $h_{ijk}$ there is an infinite 
set of triangles with areas $A_{ijk}\left( l_1,l_2,l_3\right)$
with $l_1,l_2,l_3 \in {\bf Z}$. 
\footnote{As we increase the values of 
$l_1,l_2,l_3$, the area of the corresponding triangle also increases 
by a factor proportional to the compactification radii,
and so, if these are not too small, the infinite sum (\ref{yukiguay})
can be estimated by a few terms of the series.}
The latter are triangles connecting
copies of the intersections in the ${\bf R}^6$ covering space of the
6-torus
\footnote{Recall that a two-torus can be defined as 
$T^2 = {\bf R}^2/\Lambda_{(2)}$, where 
$\Lambda_{\vec{a}, \vec{b}} = 
\{n \vec{a} + m \vec{b} \ \arrowvert \ n, m \in \int \}$ 
is a two-dimensional lattice that identifies points in ${\bf R}^2$
under translations $\vec{a}$, $\vec{b}$. 
This definition can be
generalized to $d$-dimensional tori as $T^d = {\bf R}^d/\Lambda_{(d)}$,
and we say that ${\bf R}^d$ is the covering space of $T^d$.}. 
The additional phases in eq.(\ref{yukiguay}) are
important. 
As discussed in more detail in \cite{cim5}, $\phi (l_1,l_2,l_3)$ are 
phases which are present in the generic case in which 
there are non-trivial Wilson lines of the $U(1)$'s associated
to the branes participating in the coupling
They are the crucial source of  CP-violation 
in the present formalism
\footnote{An additional source of complex phases may come 
from the inclusion of a background antisymmetric B-field,
i.e., a complexification of the tori radii. We will not 
consider this possibility here.}.

%

Let us now compute the Yukawa couplings for the specific
MSSM-like configuration described in the previous section.
For simplicity, we will suppose that branes $b$ and $c$ are one on top
of each other in the first torus and that their Wilson line backgrounds
yield equal phases in such torus. This will set the $\mu$-term to zero.
In this model the up-quark Yukawas correspond to triplets of 
intersections between branes $a$, $b$ and $c$ (yielding $abc$ 
triangles) whereas those of down-like quarks correspond to 
intersections of $a$, $b$ and $c$* ($abc$* triangles).
At the intersection of branes $b$ and $c$, $c$* lie the Higgs 
fields. The left-handed quarks appear at the three intersections 
of $a$ and $b$ (labelled by $i = 0, 1, -1$) whereas the right-handed 
up(down)-quarks (labelled by $j$ ($j$*) $= 0, 1, -1$) live 
at the $ac$ ($ac$*) intersections. The Yukawa couplings are thus labelled 
$h_{ij}^U$, $h_{ij*}^D$, and depend on the areas $A_{ij}^U(l_1,l_2,l_3)$,
$A_{ij*}^D(l_1,l_2,l_3)$. In fact, in the present case it turns out 
that all the relevant geometry takes place in the 2$^{nd}$ and 3$^{rd}$ 
torus and there is no instanton sum over $l_1$.
After some algebra one finds that the triangular areas are 
 given by
\footnote{The formula  (\ref{areaguay3}) has been computed for the 
specific model of table \ref{MSSMwrappings} with $\rho = 1$. It turns, 
out, however, that similar formulas are obtained for the case 
$\rho = 1/3$, under the interchange $R_1^{(2)} \leftrightarrow R_2^{(2)}$,
$R_1^{(3)} \leftrightarrow R_2^{(3)}$. Hence, at least for the purposes 
of this paper, both models can be considered to give the same physics.}
 \cite{cim5}
\beq
A_{ij}\left(l_2,l_3 \right) = \oh (2\pi)^2
\left(\left( R^{(2)} r_2 \right)^2 +
\left( 3 \chi R^{(3)} r_3 \right)^2\right)^\oh
\left(\left( R^{(3)} r_3 \right)^2 +
\left( 3 \chi R^{(2)} r_2 \right)^2\right)^\oh,
\label{areaguay3}
\eeq
where we have omitted the radii subindices by defining 
$R^{(\a)} = R_1^{(\a)}$. Recall also that we have 
$\chi = R_2^{(2)} /R_1^{(2)}=R_2^{(3)}/R_1^{(3)}$. The 
$r^{\alpha }$ and $\phi^{\alpha}$ quantities for $abc$ (U-quark)  
and $abc$*) (D-quark) Yukawas are defined in table \ref{param}.
\begin{table}[htb]
\renewcommand{\arraystretch}{1.25}
\begin{center}
\begin{tabular}{|c||c|c|}
\hline
 & $abc$ triangle & $abc$* triangle \\
\hline
\hline
$r_2$ & $\frac i3 + \eps^{(2)} + l_2$ & $\frac i3 + \eps^{(2)} + l_2$ \\
\hline
$r_3$ & $\frac j3 + \eps^{(3)} +
\tilde\eps^{(3)} + l_3$ & 
$\frac {j^*}{3} + \eps^{(3)} - \tilde\eps^{(3)} + l_3$
\\
\hline
\hline
$\phi^2$ & $\th^{(2)}$ & $\th^{(2)}$ \\
\hline
$\phi^3$ & $\th^{(3)} + \tilde\th^{(3)}$ & $\th^{(3)} - \tilde\th^{(3)}$
\\
\hline
\end{tabular}
\caption{\small Parameters in the MSSM-like model of table 
\ref{MSSMwrappings}.
\label{param}}
\end{center}
\end{table}

Here $i$ and $j$, $j$* (with values -1, 0, 1) label the intersections
in the second and third tori, respectively, 
whereas $l_\a \in \int$, $\a = 2,3$, index the infinite family
of instantons that contribute to the same Yukawa.
The positions of the branes for quarks are parametrized 
by the continuous parameters $\epsilon^{(2)}, \epsilon^{(3)}$ 
and ${\tilde \epsilon }^{(3)}$,
their precise geometrical meaning being shown in figure \ref{guay}.
When considering leptons one has to add two more of these
parameters $\epsilon_l ^{(2)}$ and $\epsilon_l ^{(3)}$ 
(since ${\tilde \epsilon }_l^{(3)} = { \tilde \epsilon }^{(3)}$)
 which have analogous geometrical meaning, exchanging  branes $a$ and $d$.

Concerning the phases in (\ref{yukiguay}) one can express $\phi$ as
\cite{cim5}
\beq  
\phi \left(l_2,l_3 \right) \ =\ \sum_{\alpha} \ r_{\alpha }\phi
^{\alpha}
\label{fases} 
\eeq
where the sum runs over the second and third  tori, $\alpha =2,3$.
The phases $\phi^{\alpha}$ are  linear combinations 
of Wilson line backgrounds of the $U(1)$'s associated to
the branes participating in the corresponding triangle.
Specifically, in the present model we have 
for the U-quark Yukawa phases \cite{cim5}
\beq
\begin{array}{rcl}
\phi^2 & = & \left( - \th_a^{(2)} - 3 \th_c^{(2)} \right)
\ \equiv \ \theta ^{(2)}  \\
\phi^3 & = & \left( \th_a^{(3)} - 3 \th_b^{(3)} \right) - \th_c^{(3)}
\ \equiv \ \theta ^{(3)} + {\tilde \theta }^{(3)}
\end{array}
\label{phis}
\eeq
For D-quarks one just replaces 
$\theta_c^{(3)} \rightarrow - \ \theta_c^{(3)}$, and hence 
${\phi^{3}} \rightarrow {\tilde \phi^{3}} \equiv 
 \theta^{(3)} - \ {\tilde \theta }^{(3)}$.
Here $\theta_{\sigma }^{(\a)}$ represents a $U(1)_{\sigma}$
Wilson line background over the $\a^{th}$ torus (see footnote \ref{Wilson}).

Note that the flavour indices $i$ for left-handed and 
$j$, $j$* for right-handed quarks are contained in $r_2$ and
$r_3$, respectively. Thus, the flavour structure 
depends on the brane configuration on the second and 
third torus only.
Eq. (\ref{areaguay3}) as well as the phases (\ref{fases})
are the basis for our phenomenological analysis in the next section.

\subsection{Symmetries and textures in the Yukawa couplings}

The Yukawa couplings so computed have certain points at which 
some  symmetries appear. 
Note to start with that the Yukawa couplings are invariant under
the replacement (see figure \ref{guay})
\beq
\epsilon ^{(2)} \ \longrightarrow \epsilon ^{(2)} \ + \  \frac {n}{3},
\ \quad \quad n \in \int
\label{simas1}
\eeq
since this just corresponds to a relabelling of the left-handed
quarks. In the same way one can also check the 
invariance of (\ref{areaguay3}), (\ref{fases}) under
\beq
\begin{array}{c}
\epsilon^{(3)} \ \longrightarrow \epsilon ^{(3)} \ + \  \frac {k}{3}
\\
{\tilde \epsilon}^{(3)} \ \longrightarrow{\tilde \epsilon}^{(3)} \ + \
\frac{\tilde k}{3}
\end{array}
\ \quad \quad k, \tilde k \in \int
\label{simas2}
\eeq
which is again equivalent to a relabelling of the right-handed quarks.
So, in full generality we could  allow $\epsilon ^{(2)}, \epsilon^{(3)}
,{\tilde \epsilon}^{(3)}$ to vary in between 0 and 1/3.

Let us now discuss certain symmetric configurations for
the Yukawa couplings in turn. To simplify the discussion,
we will define the new parameters 
$\d^{(3)} \equiv \eps^{(3)} + \tilde \eps^{(3)}$ and 
$\tilde \d^{(3)} \equiv \eps^{(3)} - \tilde \eps^{(3)}$, that
appear in the expression involving the Yukawas of U-quarks and 
D-quarks, respectively. We will also take them to range in $[0, 1)$,
although that by the invariances (\ref{simas1}) and (\ref{simas2}) 
we can always translate any value to one in the interval
$[0, 1/3)$. Thus, e.g., $1/2$ is equivalent to $\pm 1/6$ and so on.

\

{\bf i)} {\it $(\epsilon ^{(2)}, \phi^{(2)}) \ = \ (0,0),\ (0,1/2),\ 
(1/2,0)$}

If branes are so located,  there are two identical 
{\it rows} in the U-quark and D-quark mass matrices,
 and hence there are one massless U-quark and one 
massless D-quark eigenvalues.
This can be seen from (\ref{areaguay3}), since for
$\epsilon^{(2)} = 0, 1/2$ one obtains that
$A_{1j}(l_2,l_3) = A_{-1j}(-l_2 - 2\epsilon ^{(2)},l_3)$,
and so we can exactly match both the areas and phases involved in the
instanton sum (\ref{yukiguay}) for $i = 1$ with the ones for
$i = -1$. 
Thus, in these points there are
accidental $SU(2)$ global symmetries acting on 
the three left-handed quark multiplets. The general pattern for the 
Yukawa texture near one of these points would have the form
\beq
\begin{array}{cc}
h^{U,D}\ =\
& \left(
\begin{array}{ccc}
  a   &  b  &  c \\
d  & e  &   f \\
a  & b &  c
\end{array}
\right) ,
\end{array}
\label{aureo1}
\eeq
where the entries are generically complex numbers.
\footnote{Strictly speaking, we find such Yukawa textures {\em up
to irrelevant phases} which do not affect any of the physical 
quantities in our construction.
For instance, if we consider $(\epsilon ^{(2)},\phi^{(2)})=(0,1/2)$, 
then we find the textures
\begin{center}
$
\begin{array}{cc}
h^{U,D}\ =\
& 
\left(
\begin{array}{ccc}
\a & 0 & 0 \\
0  & 1 & 0 \\
0  & 0 & \bar \a
\end{array} \nonumber
\right) \cdot
\left(
\begin{array}{ccc}
a  & b & c \\
d  & e & f \\
a  & b & c
\end{array}
\right),
\end{array}
\quad \a = e^{2\pi i/3},
$
\end{center}
which lead to exactly the same quark masses and $V_{CKM}$ as 
(\ref{aureo1}).
}
It is obvious that there is, at least, one massless eigenstate 
from the presence of these two identical rows. 

\

{\bf ii)} {\it  $(\delta^{(3)}, \ \phi^{(3)}) , \ 
({\tilde \delta}^{(3)}, \ {\tilde \phi}^{(3)}) 
\ = \ (0,0),\ (0,1/2),\ (1/2,0)$}

This is somewhat analogous to the previous case.
Under these conditions two {\it columns } of
the U-quark and/or D-quark mass matrices are identical. Thus 
in this case there is an accidental $SU(2)$ 
global symmetry acting on the right-handed 
U-quarks and D-quarks respectively. 
The mass matrix will thus have the form
\beq
\begin{array}{cc}
h^{U,D}\ =\
& \left(
\begin{array}{ccc}
  \alpha    &  \rho  &  \alpha  \\
\beta   & \sigma   &   \beta  \\
\gamma   & \tau  &  \gamma
\end{array}
\right)  \ .
\end{array}
\label{aureo2}
\eeq
Again, there is at least one massless eigenstate from the presence of
two identical columns.


\

{\bf iii)} {\it  $(\epsilon^{(2)},\ \phi^{(2)}) = (1/2,1/2)$ }

In this case one can check that while 
$A_{0j}(l_2,l_3) = A_{0j}(-l_2-1,l_3)$,
signalling the presence of two instantonic contributions 
of equal magnitude in the same Yukawa coupling, the corresponding 
phases associated are opposite in sign. Thus, the total
sum of (\ref{yukiguay}), when $i = 0$, can be reordered 
in pairs of two terms that cancel each other, yielding a 
{\it vanishing row} in both the U-quark and D-quark mass matrices
\footnote{This is particularly clear in the case with 
$\chi=1/3$ mentioned below. In that case the Yukawa 
couplings of a row are proportional to the
Jacobi theta function $\theta[1/2,1/2]$, which is known
to vanish identically.}. 
In the same manner as in {\it i)}, one can also match instanton
contributions from rows $i=1$ and $i=-1$, now with a relative 
minus sign arising from the phase $\phi^{(2)} = 1/2$. This 
implies the Yukawa texture
\beq
\begin{array}{cc}
h^{U,D}\ =\
& \left(
\begin{array}{ccc}
  a   &  b  &  c \\
0  & 0  &   0 \\
-a  & -b &  -c
\end{array}
\right)  \ .
\end{array}
\label{aureo3}
\eeq
Where again $a, b, c \in \C$. Due to this very simple structure
one obtains two massless eigenstates.
  
\

{\bf iv)} {\it  $(\delta^{(3)},\ \phi^{(3)})
, ({\tilde \delta}^{(3)},\ {\tilde \phi}^{(3)}) = (1/2,1/2)$}

This is like the above case but case but for columns rather than rows.
One thus gets a texture of the form
\beq
\begin{array}{cc}
h^{U,D}\ =\
& \left(
\begin{array}{ccc}
  \alpha    &  0  & -\alpha  \\
\beta   & 0   &  -\beta  \\
\gamma   & 0  &  -\gamma
\end{array}
\right)  \ .
\end{array}
\label{aureo4}
\eeq
Again, one has two massless eigenstates.
Analogous results are obtained for the leptonic matrices.
Note that one may be close to several of this symmetry 
points simultaneously. Thus, if one is close to e.g. 
$({\tilde \delta}^{(3)},\ {\tilde \phi}^{(3)})=
(\epsilon^{(2)}, \ \phi ^{(2)}) = (1/2,1/2)$,
one would have a texture for the D-quarks close to a pattern
\beq
\begin{array}{cc}
h^{D}\ =\
& \left(
\begin{array}{ccc}
  \alpha    &  0  & -\alpha  \\
0   & 0   &   0  \\
-\alpha   & 0  &  \alpha
\end{array}    
\right) \ .
\end{array}
\label{aureo5}
\eeq

All these previous symmetric configurations correspond 
geometrically to several triangles getting the same area.
This is translated, after summing all the instantonic 
contributions with their corresponding phases into some discrete
symmetries of the mass matrices of the quarks, that 
for the above four cases may be summarized as
\begin{center}
$\begin{array}{l}
{\bf i)\ } h_{-i,j} = h_{i,j}, \\
{\bf ii)\ } h_{i,-j} = h_{i,j}, \\
{\bf iii)\ } h_{-i,j} = - h_{i,j}, \\
{\bf iv)\ } h_{i,-j} = - h_{i,j},
\end{array}
\quad \quad i,j = -1,0,1$
\end{center}
and whose combined effect yields one or two massless eigenvalues.
Experimentally, we know that the up and down quarks have very small
masses compared to the rest of the quarks. Thus
in our numerical search in the next section it turns
out that the branes will have often the tendency to sit relatively close
to some of these symmetry points in order to reproduce the data.
However, it turns out that, whenever the brane configuration
sits simultaneously in two symmetry points, no complex phases arise
in $V_{CKM}$ and thus there is no CP violation (this can be readily seen in
the case $({\tilde \delta}^{(3)},\ {\tilde \phi}^{(3)})=
(\epsilon^{(2)}, \ \phi ^{(2)}) = (0,0)$, where all the entries 
in (\ref{aureo1}) are real numbers).
Note also that the massless modes become massive once the
phases become complex, i.e., for
$\phi^{(2)}, \phi^{(3)}, \tilde\phi^{(3)} \not= 0, 1 /2$.
Let us, for instance, set $\eps^{(2)} = 0$.
Although we still have equal areas, that is
$A_{1j}(l_2,l_3) = A_{-1j}(-l_2 - 2\epsilon^{(2)},l_3)$,
the associated phases are now not equal (or opposite).
This implies that the Yukawa couplings of rows $i=1$ and $i=-1$
are an infinite sum of complex numbers with the same moduli 
but different phases. The interference pattern on the infinite
sum of instantonic contributions in \ref{yukiguay} is, thus, 
different for rows $i=1$ and $i=-1$, and the final Yukawas
are in general also different complex numbers, 
both in modulus and phase.

\

{\bf v)} {\it  The point $\chi = 1/3$}

The point  $R_2^{(2)}
/R_1^{(2)}=R_2^{(3)}/R_1^{(3)} = \chi=1/3$ is somewhat special.
Geometrically what happens  is that
branes $a$, $d$ form angles $\pm \pi /4$ with the orientifold
plane (branes $b$,$c$ form angles $\pm \pi /2$).
For $\chi = 1/3$ the Yukawa couplings take a particularly simple
form. Indeed in that case one has
\beq
A\left(r_{2}, r_{3}\right) = \oh (2\pi)^2
\left[
\left( R^{(2)} r_2 \right)^2 + \left(R^{(3)} r_3 \right)^2
\right].
\label{areaguay5}
\eeq
Now the expression is quadratic and the whole expression for the
Yukawa coupling (\ref{yukiguay}) allows us to express them
in terms of a product of Jacobi theta functions with 
characteristics, namely
\beq
h_{ij}^{abc} \ = \
\vt \left[
\begin{array}{c}
 i/3+\eps^{(2)} \\ \th^{(2)}
\end{array}
\right] (t_2) \times
\vt \left[
\begin{array}{c}
 j/3+\eps^{(3)} + \tilde\eps^{(3)} \\ \th^{(3)} + \tilde\th^{(3)}
\end{array}
\right] (t_3)
\eeq
where $t_2 = \left(R^{(2)} \right)^2 / \alpha'$ and 
$t_3 = \left(R^{(3)}\right)^2 / \alpha'$, and we have
imposed the data for $r_\a$, $\th^{(\a)}$ from table \ref{param}.
The Yukawas for $abc$* are similar, but for the substitution
$(\tilde\eps^{(3)},\tilde\th^{(3)}) \mapsto
(-\tilde\eps^{(3)},-\tilde\th^{(3)})$, and $j \mapsto j$*. 
The above theta functions are defined by
\beq
\vt \left[
\begin{array}{c}
\d \\ \phi
\end{array}
\right] (\nu,\tau) = \sum_{l \in \int}
e^{\pi i (\d + l)^2 \tau} \ e^{2\pi i (\d + l)(\nu + \phi)}.   
\label{thetacpx}
\eeq
It is easy to show that, irrespective of the
brane locations,  for $\chi =1/3$  Yukawa matrices
all have one massive eigenvalue and two massless ones.
Although the expressions for the Yukawas are quite elegant,
in our numerical search in the next section we have not found
 parameter choices with $\chi =1/3$ leading to a good description
of the experimental results, although our search has not been
exhaustive.

Let us now briefly  comment on some intriguing properties of the 
complex phases.
One would expect that all physical quantities like
mass eigenvalues and mixings should be invariant under 
an integer shift in the Wilson line phases
\beq
\phi ^\alpha \ \longrightarrow \ \phi ^\alpha \ +\ n,  \  
\quad \quad  n \in \int.
\label{shift}
\eeq
This is not obvious from (\ref{fases}), since 
the coefficients $r^{\alpha }$ are in general fractional.
This implies that {\em Yukawa matrices} $h_{ij}^U$, $h_{ij*}^D$
{\em are not invariant} under (\ref{shift}), which is quite puzzling.
One can, however, show that this invariance is indeed present
in the {\em measurable} physical quantities, that is, 
the mass eigenvalues and the CKM mixing matrix \cite{cim5}.

One can also convince oneself of the following result 
concerning phases and CP violation. In order to get a
non-negligible contribution to CP-phases, more than the first  term should
be  present in the worldsheet instanton sum in eq.(\ref{yukiguay}).
If there is only the first term (i.e. $l_1, l_2, l_3 = 0$), then
one can always  reabsorb the phases coming from Wilson lines 
into quark states, so that
the mass matrices are real. Note in particular that, if the radii are
large, then the leading contribution to Yukawa couplings will
come from the first term only, the following terms being
exponentially suppressed. As a consequence
CP-violation
 (i.e. the Jarlskog invariant) will be very small for large radii.
Indeed this turns out to be the case in the numerical analysis
in the next section. Thus we will need 
relatively small radii (so that several terms contribute 
in the instanton sum) in order to obtain
sufficient CP-violation.

Let us finish this section by making a remark about the 
absolute size of Yukawa couplings in this setting.
The above computation corresponds to the classical contribution
$exp(-S_{cl})$ to the actual Yukawa correlators. In addition
there are quantum world-sheet corrections which correspond to
quantum  fluctuations \cite{hv} 
around the flat triangle worldsheets.
Those are expected to be flavour-independent and to affect
in a similar way both U- and D-quark Yukawas. Thus the
actual Yukawa couplings will have a general form \cite{hv}
$Y_{ij}^{U,D} = h_{qu} h_{ij}^{U,D}$, with $h_{qu}$ including the
quantum fluctuation factor and $h_{ij}^{U,D}$ being the classical
contribution discussed above.

Furthermore, in order to make contact with the explicit Yukawa 
coupling constants
appearing in the Standard Model, all gauge and fermion fields should have 
kinetic terms normalized to one. This gives an additional factor 
which, e.g., in the heterotic case is proportional to the 
gauge coupling constant $g$. In the present case there is no
unified gauge coupling but different couplings for each gauge group. 
So in our case one expects some geometrical average $g_0$ of the 
gauge couplings. Finally, if we want to compare with 
fermion data at say the $Z^0$-scale, there will be in general
a loop running from the string scale, at which all the above 
expressions apply, down to the $Z^0$-scale. 
Combining  these two factors  with the quantum
fluctuation factor $h_{qu}$ 
one has for the Yukawas at the $Z^0$ scale
\beq
Y_{ij}(M_Z)\ =\   \xi (M_Z)  h_{qu}g_0 h_{ij} \ \equiv \ h_0
h_{ij}
\label{fudge}
\eeq
where $\xi (M_Z)$ is a renormalization group factor.
Those corrections are expected to be similar  for 
both U- and D-quarks, since the leading effects should
come from QCD loops. Thus we will take $h_0^U = h_0^D \equiv h_0$.
On the other hand, the running effect on leptons should be
much smaller and one expects approximately to
have $h_0^L =h_0/\xi(M_Z)$.
In any event, in the numerical computation we will leave $h_0$
and $h_0^L$ as free parameters.

\section{Reproducing the experimental results}

Let us now see if the above brane configuration is able to
reproduce the quark spectrum, mixing angles and CP violation.
We will also consider the charged lepton  and Dirac 
neutrino masses below. 
The Yukawa couplings we computed in the previous section
applied at the string scale. If the string scale is high 
 the running effects for the quarks down to the 
weak scale may be
large. As we mentioned above we will reabsorb those effects in the
definition of $h_0^{U,D}\equiv h_0$ and $h_0^L$ and then
we will compare with data at the  $Z^0$  scale.

As we said, this model can also be
understood as a Pati-Salam model in which brane 
separation (adjoint Higgsing) breaks it to
$SU(3)\times SU(2)_L\times U(1)^2$. In the presence of 
exact Pati-Salam symmetry and a minimal
Higgs sector one has (before r.g. running)
\beq 
 M_U\ =\ M_{\nu} \ =\ tg\beta \ M_{D}\ =tg\beta \ M_{L}
\label{massps}
\eeq
where $tg\beta = \frac {<{\bar H}>}{<{ H}>}$.
This leads to identical hierarchies for all fermions
and no CKM, so this cannot give, as it stands,  a
good description of the SM fermion spectrum.
However this may be a good starting point since
after all the experimental hierarchical structure is somewhat 
similar for the different quarks and leptons 
and the CKM mixing matrix is not far away from unity.
So it seems it could be a good idea to perturb around
such a D-brane configuration.
We now separate from this PS symmetry point
by setting 
$\epsilon _l^{(2)}\not= \epsilon ^{(2)}$ and
$\epsilon _l^{(3)}\not= \epsilon ^{(3)}$. Then 
the $SU(4)$ breaks to $SU(3)\times U(1)$.
In addition we set ${\tilde \epsilon}^{(3)}\not= 0,1/6$ and
then $SU(2)_R$ breaks to $U(1)_c$. 
Naively, our field theory experience would tell us that this
is not going to affect the electroweak Yukawa couplings,
those are different couplings which have nothing to do 
with the adjoints. However, the implementation in terms of
wrapping D-branes tells us that the Yukawa couplings are 
affected by this breaking: since the relative brane locations vary,
the areas of the triangles also vary and the equations
(\ref{massps}) no longer hold.
\footnote{From the field theory point of view this means that the
non-renormalizable operators of type $\psi_R\psi_LH(adjoint)^n$
are playing a role. The difference is that in our D-brane
realization all these non-renormalizable terms are automatically 
summed to all orders 
 by considering how the area of the triangles
involved in the Yukawa computation have changed.
Note that the $\epsilon $'s and $\phi $'s 
parametrizing brane locations and Wilson lines
correspond to vev's for complex adjoint scalar fields in 
these brane settings. Thus, e.g., an expansion on small separations 
$\epsilon$'s from the PS limit correspond to an expansion 
involving powers of adjoint scalars.}

Once we have separated from the PS symmetry point,
the mass matrices for U- and D-quarks are no longer 
proportional and are given by
\beq
M^U_{ij}\ =\ h_0\ h^U_{ij}<{\bar H}> \ ;\ 
M^D_{ij}\ =\  cotg(\beta) h_0\ h^D_{ij}<{ \bar H}>
\label{massesyuk}
\eeq
respectively, where
\beq  
 \ \sqrt{|<{\bar H}>|^2+
|<{H}>|^2 }\ =\ \frac {\sqrt{2}M_W}{g_L} \ =\ 174 GeV
\label{mw}
\eeq
and $h^U_{ij}$, $h^D_{ij}$ are taken from eqs.
(\ref{yukiguay}), (\ref{areaguay3}), (\ref{fases})
and the data on table \ref{param}.
They will explicitly depend on the geometry of the
tori ($R^{(2)},R^{(3)}, \chi $),  the $\epsilon $'s
which parametrize the brane positions and the phases $\th$
 coming from the Wilson lines.
The mass matrices are in general not symmetric and becomes
convenient to work with the hermitic matrices 
$M^U(M^U)^\dag$ and $M^D(M^D)^\dag$ in order to obtain the 
mass eigenvalues and the unitary matrices $U_L^U$, $U_L^D$ 
\beq
\begin{array}{c}
(U_L^U)^\dag \left(M^U(M^U)^\dag\right) U_L^U \ =\ 
\left(
\begin{array}{ccc}
m_u^2  &  0 &  0   \\
 0 & m_c^2 & 0\\
0 & 0 & m_t^2  
\end{array}
 \right) \\
(U_L^D)^\dag \left(M^D(M^D)^\dag\right) U_L^D \ =\
\left(
\begin{array}{ccc}
m_d^2  &  0 &  0   \\
 0 & m_s^2 & 0\\
0 & 0 & m_b^2
\end{array}
 \right)
\end{array} 
\label{diag}
\eeq
which diagonalize them.
Then the CKM mixing matrix will be given by
\beq
V_{CKM} \ =\  (U_L^U)^\dag\ U_L^D
\label{ckm}
\eeq
and will generically be complex. In the present case it is 
convenient to express CP-violation in a convention independent way,
i.e., in terms of the Jarlskog invariant J \cite{jarlskog} which may be
computed in terms of four complex entries of $V_{CKM}$,
for example:
\beq
J\ =\ Im(V_{cs}V_{us}^*V_{ud}V_{cd}^*)
\label{jarlskog}  \ .
\eeq
In terms of the `standard parametrization' \cite{pdg}, this may be
expressed in terms of three real mixing angles $\theta _{ij}$ and the 
complex phase $\delta _{13}$ :
\beq
J\ = \ s_{12}s_{13}s_{23}c_{12}c_{23}c_{13}^2 sin(\delta _{13})
\label{d13}
\eeq
where $s_{ij}=sin(\theta _{ij})$, $c_{ij}=cos(\theta _{ij})$.

The mass matrices depend on the following 8 real parameters:
$h_0$, $tg\beta$, $R^{(2)}$, $R^{(3)}$, $\chi$,
$\epsilon ^{(2)}$, $\epsilon^{(3)}$ and 
${\tilde \epsilon }^{(3)}$. In addition there are the phases 
$\theta ^{(2)}$ , $\theta ^{(3)}$ and ${\tilde \theta}^{(3)}$.
We have made a computer search for choices of these parameters 
able to describe the observed spectrum of quark masses,
mixing angles, CP-violation phase and charged lepton masses.
We will discuss first the quark sector.
 We find different regions in parameter space in which
all those data can be reasonably adjusted and we show some possible
choices in table \ref{modelos} and the corresponding 
results in table \ref{resultados}.
Note that the 
 values for $h_0$ and $tg\beta $ are fixed by the condition that 
they reproduce the results $m_{b}(M_{Z^0})=3.0$ GeV,
$m_t(M_{Z^0})=180 $ GeV. 

\begin{table}[htb] \footnotesize
\renewcommand{\arraystretch}{1.25}
\begin{center}
\begin{tabular}{|c|c|c|c|c|c|c|c|c|c|c|c|}
\hline
 Model  & $h_0$  &  $ tg\beta $   &  $R^{(2)}$  &  $R^{(3)}$ & $\chi $
 & $\epsilon ^{(2)}$  &  $\delta ^{(3)}$ & ${\tilde \delta} ^{(3)}$   
& $\phi ^{(2)}$  &  $\phi ^{(3)}$ & ${\tilde \phi} ^{(3)}$
\\
\hline\hline
 A   &  1.169  &  33.33 & 0.105  & 0.66  & 2.552  & 0.09 &
0.50056 &  0.03 & 0.43 & 0.43 & 0.485  \\
\hline
B  &  0.715  &  74.48 & 0.245  & 0.17  & 1.71 &  0.024&
0.5004 &  0.041 & 0.1405& 0.472 & 0.384  \\
\hline
C  &  0.715  &  74.20 & 0.2516  & 0.1714  & 1.71  & 0.024 &
0.50059 &  0.029 & 0.14 & 0.47383 & 0.386 \\
\hline
D  &  0.493  &  11.04 & 0.1  & 0.15  & 1.75  & 0.452 &
0.0011 &  0.5091 & 0.4539 & 0.0 & 0.25 \\
\hline
E  &  0.597  &  63.7  & 0.3024   & 0.2  & 1.75  & 0.186 &
0.0009 &  0.1254 & 0.098 & 0.50001 & 0.403 \\
\hline \end{tabular}
\end{center} \caption{ Some choices of parameters
leading to the results shown in table \ref{resultados}.
The radii are given in units of $M_s^{-1}$.}
\label{modelos}
\end{table}

\begin{table}[htb] \footnotesize
\renewcommand{\arraystretch}{1.25}
\begin{center}
\begin{tabular}{|c|c|c|c|c|c|c|c|c|c|c|}
\hline
 Model  & $m_u$ &  $m_d$   &  $m_s $  &  $m_c$ & $m_b$
 & $m_t$  &  $|V_{12}|$ & $|V_{23}|$
& $|V_{13}|$  &   J
\\
\hline\hline
 A   & 1.28   &  4.32  & 81.75   & 669.1  & 3.0  & 180.0 &
0.224 &  0.0499 & 0.0023    &  $0.9 \times 10^{-5}$  \\
\hline
 B   & 4.37   &  6.90  & 65.73   & 576.46  & 3.0  & 180.0 &
0.2194 &  0.0386 & 0.0049    &  $ 3.0 \times 10^{-5}$  \\
\hline
 C   & 2.81    &  7.90   & 71.07   & 646.97  & 3.0  & 180.0 &
0.2259 &  0.0408 & 0.0048    &  $2.88 \times 10^{-5}$  \\
\hline
 D   & 0.95    &  1.97   & 73.03   & 607.88  & 3.0  & 180.0 &
0.2199 &  0.0484 & 0.0025    &  $2.61 \times 10^{-5}$  \\
\hline
 E   & 0.91    &  1.04    & 104.05   & 709.57   & 3.0  & 180.0 &
0.225 &  0.0381 & 0.0058    &  $1.14 \times 10^{-5}$  \\
\hline \end{tabular}
\end{center} \caption{ Quark masses, mixing angles and
Jarlskog invariant at the $Z^0$ scale
for the model parameters of table \ref{modelos}.
The  masses for the two heaviest quarks are given in GeV
and the rest in MeV.}
\label{resultados}
\end{table}

Let us summarize our findings on a few points:

\begin{itemize}

\item
Appropriate results are obtained if the branes are close to
some symmetry points as discussed in the previous section.
We have found all the acceptable results to be very close
to points with 
%
\beq
\delta^{(3)}  \ = \  \phi ^{(3)} \ =\ 0,\ 1/2 \ 
\label{aureo}
\eeq
with a good accuracy. Also some examples 
are in addition close to other symmetry points like 
$\epsilon^{(2)}  \ = \  \phi ^{(2)} \ =\ 1/2$
(example D) .
 It is easy to understand the origin of
eq.(\ref{aureo}). The biggest intergeneration hierarchy in the standard 
model is that of $m_{up}/m_{top}\propto 10^{-5}$ (compared to the 
D-quark hierarchy which is $m_{d}/m_{b}\propto 10^{-3}$ only).
If the brane $bç$ is close to a symmetry point $\delta^{(3)}=0,1/2$
in the third torus, then we mentioned in the previous section that
a massless U-quark eigenvalue is obtained, explaining the 
presence of the U-quark hierarchies. On the other hand,
the hierarchy of the D-quarks is milder and is obtained
by having, e.g., either ${\tilde \delta }^{(3)}$ 
 or/and $\epsilon^{(2)} \simeq 0,1/2$.

It is interesting to show the type of {\it texture } for
the U- and D-quark masses that one obtains. For example for the 
choice of parameters D in table \ref{modelos} one finds
for the moduli of the entries $|Y^{U,D}_{ij}|$ and the
phases $\sigma ^{U,D}_{ij}$ (in radians) of Yukawa couplings 
\beq
|Y^U|   =
\left(
\begin{array}{ccc}
0.958  &  0.843 &  0.959   \\
0.245  & 0.221 &  0.246 \\
0.783 & 0.697  &  0.784
\end{array}
 \right), \
\sigma ^U  =
\left(
\begin{array}{ccc}
-0.290  &  -0.288 &   -0.290   \\
 -0.455  &  -0.453 & -0.455\\
 0.504  & 0.502 & 0.504
\end{array}
 \right).   
\label{textureu}
\eeq
\beq
|Y^D|   =   
\left(
\begin{array}{ccc}
0.185  &  0.087 &  0.195   \\
0.056  & 0.032 &  0.058 \\
0.162  & 0.086  &  0.170
\end{array}
 \right), \
\sigma ^D  =
\left(
\begin{array}{ccc}
-0.494    &  -0.252 &   -0.086   \\
-0.614   & -0.454  & -0.282\\
 0.300   & 0.509  &  0.669
\end{array}
 \right).
\label{textured}
\eeq
One can qualitatively understand this structure by noting
that the brane location parameters D are relatively close
to a symmetry point with 
\beqa
\epsilon^{(2)} & = \ \phi ^{(2)} & =\ 1/2 
\label{tachan1} \\ 
\delta^{(3)} & = \ \phi ^{(3)} &  =\  0 
\label{tachan2} \\ 
{\tilde \delta}^{(3)} & =\ 1/2  & 
\label{tachan3}
\eeqa
and $\tilde \phi^{(3)}$ at some intermediate value
between $0$ and $1/2$ (namely $\tilde \phi^{(3)} = 1/4$).
The brane configuration in the 2$^{nd}$ and 3$^{rd}$ tori 
corresponding to this symmetric configuration is depicted in 
fig.\ref{gold}.
%
\begin{figure}[t]
\centering
\epsfxsize=5.5in
\epsffile{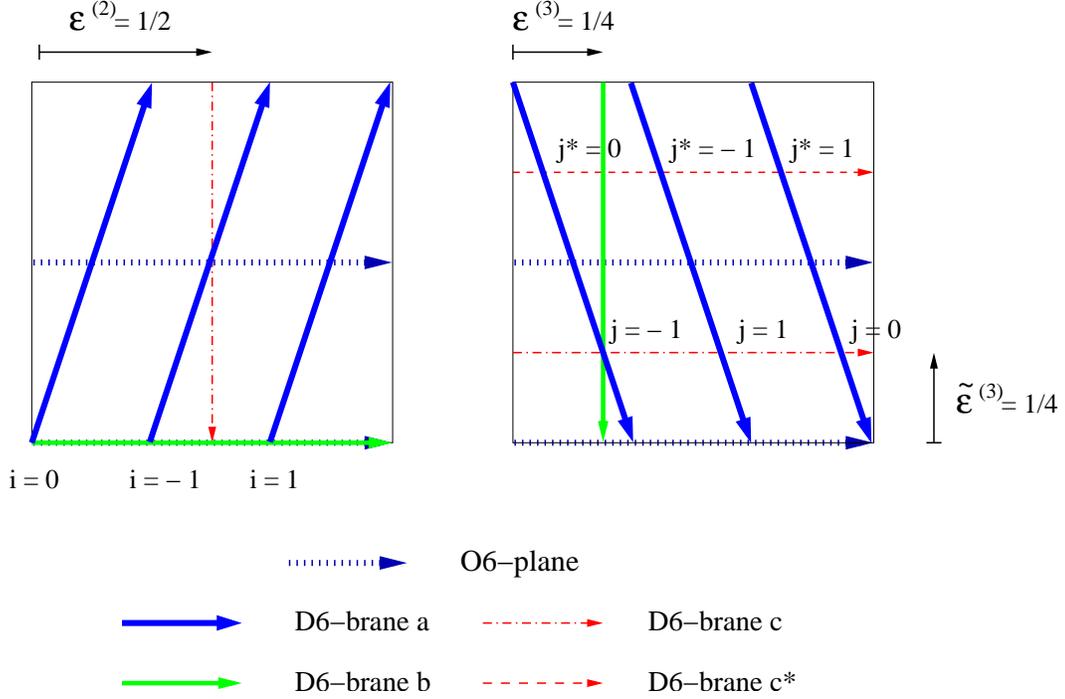}
\caption{\small{Brane configuration corresponding to the 
symmetry point in (\ref{tachan1}), (\ref{tachan2}) and
(\ref{tachan3}). For concreteness, we 
only show the 2$^{nd}$ and 3$^{rd}$ tori.}}
\label{gold}
\end{figure}
%
According to our discussion on symmetries and textures in
previous section, due to the proximity to the condition
(\ref{tachan1}) both textures of U-quark and D-quark
Yukawas should present a behaviour quite similar to
(\ref{aureo3}), with the entries of one row much 
smaller than the other two.
In addition, due to (\ref{tachan2}), the U-quark matrix
should have two equal rows, as in (\ref{aureo2}), whereas
the D-quark matrix should have some intermediate behaviour 
between (\ref{aureo2}) and (\ref{aureo4}).
Indeed, the mass matrices in (\ref{textured}) behave 
qualitatively like that.
 So the example D in table \ref{modelos} 
is able to reproduce the experimental results due to its
proximity to points with special symmetries. 
Something analogous happens with the other choices of
parameters in table \ref{modelos}, being all relatively close 
to some symmetry point
\footnote{They cannot, however, sit {\it exactly} on such points,
since this would imply no CP violation.}.

\item  The brane configurations able to reproduce the data  
may be understood as a continuous  deformation of a left-right
symmetric configuration. As we said, for ${\tilde \epsilon}^{(3)}=0$
(or, equivalently, $\delta ^{(3)}={\tilde \delta }^{(3)}$)
the brane $c$ is on top of its mirror $c$* and there is a 
gauge enhancing $U(1)_c\rightarrow SU(2)_R$ to 
a left-right symmetric model. Under those circumstances 
the U-quark and D-quark mass matrices are proportional 
and hence there is no CKM mixing. As we switch on
${\tilde \epsilon}^{(3)}\not= 0$ $M_U$ and $M_D$ cease to be 
proportional and CKM mixing is generated
\footnote{One can measure how strongly $SU(2)_R$ is broken
in terms of $2{\tilde \epsilon}^{(3)} = \delta^{(3)}-
{\tilde \delta}^{(3)}$, which is the distance
between 
branes $c$ and $c^*$ in the third torus as well as the 
Wilson line background ${\tilde \th}^{(3)} = 
\th^{(3)} - \tilde\th^{(3)}$ breaking $SU(2)_R$. 
In fact the mass of the $W_R$ gauge boson 
may be written in terms of those two parameters. 
Going through the examples of the tables one finds that 
$M_{W_R}$ is always of order the string scale. So one can conclude that
the underlying $SU(2)_R$ symmetry is strongly broken.}.  
This may be numerically seen in fig. \ref{cruce}.

\begin{figure}[ht]
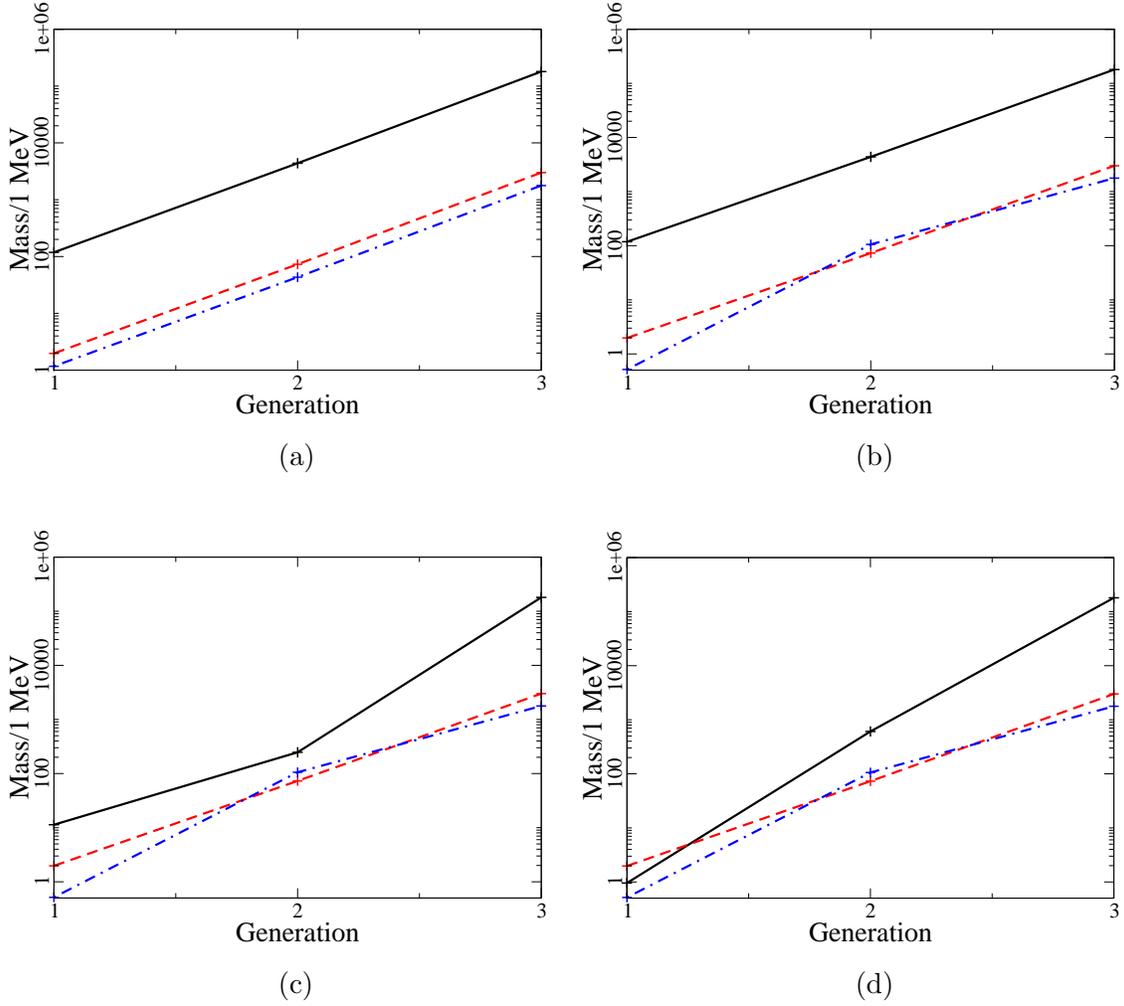

\begin{center}
\begin{tabular}{ll}
\\
\hskip -0.5cm
\epsfig{file=patisalam.eps, height=5.5cm} &
\epsfig{file=LR_sym.eps, height=5.5cm}\\
\hskip 3.1truecm
{\small (a)}            &
\hskip 3.65truecm
{\small (b)}\\
\\
\hskip -0.5cm
\epsfig{file=wilson.eps, height=5.5cm} &
\epsfig{file=experimental.eps, height=5.5cm}\\
\hskip 3.1truecm
{\small (c)}            &
\hskip 3.65truecm
{\small (d)}\\
\end{tabular}
\end{center}
\caption{\small{
The masses of U-quarks (solid line), D-quarks (dashed line)
and charged leptons (dashed-dotted line)
at the $Z^0$ scale
for the three generations.
Figure (d) shows the results for the choice of parameters
D in tables \ref{modelos} and \ref{modelosl}, in good agreement
with experiment. Figures (a), (b) and (c) show how that
brane configuration may be continuously reached from
an underlying Pati-Salam configuration by shifting of
branes and Wilson line breaking (see text).}}
\label{cruce}
\end{figure}
%
In fig.\ref{cruce}(d), the solid and dashed lines 
show the masses of U- and D-quarks masses respectively 
of the three generations as reproduced by model D in table 
\ref{modelos}. One can restore $SU(2)_R$ symmetry 
by setting $\delta ^{(3)} ={\tilde \delta}^{(3)}$ 
(fig.\ref{cruce}(c)) and setting 
$\phi ^{(3)} ={\tilde \phi}^{(3)}$  (fig.\ref{cruce}(b),(a)).
Then one observes that the masses of U-quarks and 
D-quarks are proportional, as it should in a
left-right symmetric model. One can also check that
in this LR-symmetric limit the mixing disappears.
Note also that one observes that 
 {\it this $SU(2)_R$ breaking is the cause for  the observed 
inverted hierarchy $m_{up}<m_{down}$ as well as the 
appearance of mixing} (and also CP-violation).

\item
One can easily obtain fermion mass hierarchies
consistent with experimental data. However,
getting
 substantial CP violation   
requires that there are several instanton numbers
 contributing significantly to
 the world-sheet instanton sum in eq.(\ref{yukiguay}).
 This happens when the torus radii  
$R^{(2)}$ and $R^{(3)}$ are smaller than $M_s^{-1}$,
\footnote{Note that this does not affect the radius
of the first torus whose size is unconstrained by
Yukawa couplings.} otherwise only the first term
contributes and CP-violation is suppressed.
Note however that the radii  $R^{(2)}$ and $R^{(3)}$
cannot be too small, since then the instanton sum
diverges and the quantum piece of the Yukawa coupling
becomes relevant. We have checked in all the numerical
examples that the corresponding instanton sum converges.
In most of the cases the terms in the sum
eq. (\ref{yukiguay}) become negligible for
$|l_2|,|l_3|\geq 6$.

\item
The complex phases coming from Wilson lines not only 
may  account for the observed CP-violation. 
In addition, they have important influence on the
values of the masses of the lightest fermions. 
Recall that, e.g., as we mentioned in the previous section
the symmetry points with $\epsilon ^{(2)},\ \delta ^{(3)},\ 
{\tilde \delta }^{(3)}\ =\ 0,\ 1/2$ give rise to massless 
fermions only if, in addition, one has real phases
$\phi^{(2)},\ \phi^{(3)},\ \tilde\phi^{(3)}\ =\ 0,\ 1/2$.  
Thus, for non-vanishing 
complex phases these massless modes will get massive.

\end{itemize}

We conclude that one can reproduce in a rather satisfactory way 
the quark masses, mixings and CP-violation in terms 
of the present model. 
Note that we have not done a fully systematic search 
for ranges of parameters able to reproduce the experimental results
and the choices shown in the tables are just possible examples. 
Some values of parameters are particularly successful,
like the choice D in table \ref{modelos}.
The values we found for $tg\beta $ vary in
the range $10-74$ and the value  for $h_0$ is of order one,
as expected from our discussion at the end of section 3.

Although there are a number of free parameters, note
that the brane location parameters $\epsilon $, $\delta $ and phases 
in almost all cases sit close to 
symmetry points. One may speculate that  those parameters
could  be fixed 
by some symmetry in the dynamics which eventually 
determines  the geometry of the brane configuration. 
One can consider an scenario in which all the branes 
are located on top of some symmetry point 
like, e.g., that in (\ref{tachan1}), (\ref{tachan2}),
(\ref{tachan3}) 
and some (e.g., loop) corrections give rise to a small deviation 
from those values. In such scenario the number of free 
parameters would be essentially reduced to the geometric moduli 
and $tg\beta $.

A similar study can be done for charged leptons and Dirac 
neutrino masses. In fact, once fixed 
$\epsilon _l^{(2)}$ and $\delta _l^{(3)}$ (plus phases)
to reproduce the
charged lepton spectrum, {\it Dirac
neutrino masses as well as Dirac mixing angles
and leptonic CP-violating phases are fixed},
since the geometry of the whole brane configuration
gets completely determined.
This is a very interesting point.  
However, in the case of neutrino masses
the structure of the 
Majorana mass for the right-handed neutrinos
is very relevant. As we
mentioned above, once some linear combinations of sneutrinos get 
a vev (which may be triggered by a FI-term), the  
$SU(3)\times SU(2)\times U(1)_c\times U(1)_{B-L}$ symmetry
is broken to the SM group. One can also see that 
in the same process r.h. neutrinos get a mass \cite{cim12}.
The latter mechanism is somewhat different to the
class of Yukawa couplings we are considering here.
Thus we will present a complete analysis of the leptonic
and neutrino sector in this class of models in a separate
publication.
Here we will contempt ourselves with  the analysis of the charged leptonic
sector. Again one can reproduce the observed spectrum of charged leptons
in terms of the location parameters of the leptonic brane $d$,
$\epsilon _l^{(2)}$ and $\epsilon _l^{(3)}$ (and phases). 
In addition, once we have adjusted the quark and charged lepton
sector, the Dirac neutrino masses and (Dirac) neutrino mixing 
matrix as well as leptonic CP-phases are fixed.
Table \ref{modelosl} shows some choices of the leptonic
parameters (corresponding to the previous quark sector choices
A-E) giving rise to the results in table \ref{resultadosl}.
Of course, only the charged lepton masses are directly measured at present,
and those can be adjusted remarkably well. As we said, the rest of the
information is certainly relevant for the structure of physical 
neutrino masses and leptonic CP phases, but a knowledge of 
the r.h. Majorana masses should be first provided.
 Let us only mention that
the leptonic CP-phases  may be large  in some cases and also that 
sometimes one of the Dirac neutrino masses may be essentially vanishing.
That happens, for instance, in model D because one sits 
on the point $(\delta _l^{(3)}, \phi_l^3)=(1/2,1/2)$,
giving a vanishing column in the Dirac neutrino
mass matrix. 
The corresponding sneutrino may then
acquire a large vev (breaking (B-L)) without giving rise to a 
large mass term of type $HL$. This large vev would then give rise
to large Majorana masses for all r.h. neutrinos,
and a standard seesaw mechanism could then be at work.

We also find that the brane configurations 
able to reproduce in addition 
the observed spectrum of charged leptons 
may be understood as a deformation of a Pati-Salam
symmetric configuration.
As we said, if  the leptonic brane sits on top
of the baryonic stack (i.e.,
$\epsilon _l^{(2)}=\epsilon ^{(2)}$ and
$\epsilon _l^{(3)}=\epsilon ^{(3)}$) the gauge symmetry
is further enhanced to a full Pati-Salam symmetry
$SU(4)\times SU(2)_L\times SU(2)_R$.
Numerically we find (modulo 1/3) 
$\epsilon_l^{(2)}-\epsilon^{(2)}$$ \simeq
\delta_l^{(3)}-\delta^{(3)} \simeq 
0.04 - 0.09$
indicating that   the brane configuration able to 
reproduce the quark masses and mixings as well as 
the charged lepton masses may be considered as a
 deformation of a brane configuration leading to a 
Pati-Salam $SU(4)$  symmetry
\footnote{Note that this applies if  the baryonic and 
leptonic branes sit on top of each other in the first torus,
which is not necesarily the case. The Yukawa couplings are 
not sensitive to the geometry in the first torus.}. 
 This is again illustrated in fig.\ref{cruce}.
In fig.\ref{cruce}(a) we have set 
$\epsilon^{(2)}=\epsilon_l^{(2)},
\epsilon^{(3)}=\epsilon_l^{(3)}$ and also equal phases.
Then the masses for D-quarks and charged leptons
become proportional. In fact in the PS limit they should 
be equal at the string scale, but recall that we 
have allowed for different original values for $h_0$ and $h_0^l$,
which could take in to account e.g., the renormalization 
group running from the string scale to the $Z^0$ scale.

\begin{table}[htb] \footnotesize
\renewcommand{\arraystretch}{1.25}
\begin{center}
\begin{tabular}{|c|c|c|c|c|c|}
\hline
 Model  & $h_l$  
 & $\epsilon_l ^{(2)}$  &  ${\tilde \delta}_l ^{(3)}$ 
& $\phi_l ^{(2)}$  &  $\phi_l ^{(3)}$ 
\\
\hline\hline
 A   &  0.684 &  0.35& 0.0011  & 0.35 & 0.3104
  \\
\hline
B  &  0.43  &  0.4 & 0.155  & 0.23  & 0.29
  \\   
\hline
C  &  1.51 &  0.17 &  0.03   & 0.47  & 0.38
  \\
\hline
D  &  0.166 &  0.499 & 0.508  & 0.1845 & 0.25
\\
\hline
E  &  0.141  &  0.0 & 0.06   & 0.06 & 0.14 \\
\hline \end{tabular}
\end{center} \caption{ Choices for the leptonic  parameters
corresponding to the ones in table \ref{modelos} and
leading to the results in table \ref{resultadosl}.}
\label{modelosl}
\end{table}

\begin{table}[htb] \footnotesize
\renewcommand{\arraystretch}{1.25}
\begin{center}
\begin{tabular}{|c|c|c|c|c|c|c|c|c|c|c|}
\hline
 Model  & $m_e$ &  $m_{\mu }$   &  $m_{\tau}$   &  $m_{\nu_e}$ & 
$m_{\nu_ \mu}$
 & $m_{\nu_\tau}$  &  $|V^l_{12}|$ & $|V^l_{23}|$
& $|V^l_{13}|$  &   $J_l$
\\
\hline\hline
 A   & 0.512   &  105.66  & 1777   & 18.15  & 3.615  & 124.77 &
0.127 &  0.098 & 0.0002    &  $1.16 \times 10^{-6}$  \\
\hline
 B   & 0.516   &  105.59  & 1777   & 309.5  & 4.683  & 83.95 &
0.052 &  0.228 & 0.021     &  $ 15.3 \times 10^{-5}$  \\
\hline
 C   & 0.513   &  105.48  & 1777   & 1.90 & 1.698  & 160.22 &
0.021 &  0.057 & 0.0018     &  $  5.7 \times 10^{-7}$  \\
\hline
 D   & 0.516    & 105.52   & 1777   &  0.0   & 0.652  & 229.45 &
0.200 &  0.0648 & 0.0232    &  $5.9 \times 10^{-7}$  \\
\hline
 E   & 0.511    & 105.56   & 1777   &  35.72   & 5.989   & 82.27 &
0.009 &  0.128  & 0.0044    &  $1.39 \times 10^{-6}$ \\ 
\hline \end{tabular}
\end{center} \caption{ Charged lepton  and Dirac neutrino masses,
 mixing angles and (leptonic) Jarlskog invariant 
at the $Z^0$ scale 
for the model parameters of
tables \ref{modelos} and \ref{modelosl}.
All masses are given in MeV except for the two heaviest (Dirac) neutrino
masses which are given in GeV.}
\label{resultadosl}
\end{table}

\section{ Final comments and conclusions}

A number of comments are in order:

\begin{itemize}

\item A few words about the particular brane configuration 
analyzed. 
For the comparison with experimental data we have considered
a new D6-brane configuration which is particularly simple.
At the local level the model has $\cn = 1$ SUSY and the MSSM 
chiral spectrum. It has the minimal Higgs sector of the MSSM
and the $\mu$-parameter is governed by both the distance between
the branes $b$ and $c$ and their relative Wilson line 
in the first torus. 
The gauge group is that of the SM plus
an extra $U(1)$ familiar from left-right symmetric models.
The latter may be broken spontaneously by a right-handed 
sneutrino vev, as explained in ref.\cite{cim12}. If this configuration
of D6-branes is wrapping a 6-torus, there are in addition 
adjoint scalars and fermions with respect to the gauge group,
which would become massive after SUSY breaking.
Let us remark that 
Type II string theory models with D-branes like this have
consistency constraints from the cancellation of the 
Ramond-Ramond (RR) tadpoles. Specifically, 
the theory has certain antisymmetric fields under which 
D-branes (and orientifold planes) are charged. Tadpole
conditions come from imposing that the overall RR-charges 
of the D-brane configuration vanishes.
In the example considered in the text the D-brane 
configuration does not cancel by itself the RR-tadpoles.
Thus  in order to cancel RR-tadpoles 
there should be further D-branes in the model.
 Since however, the
SM brane configuration is anomaly-free, the extra branes will have
vanishing intersection numbers with the SM branes, i.e., they will not
give rise to extra chiral massless modes, being some sort
of `hidden sector' in the model \cite{cim12}.  
One can also check that in the simple toroidal
case the complete D-brane configuration including the SM branes
and the extra D-branes is necessarily non-SUSY \cite{bgkl,cim12}. 
Thus one would be forced to lower the string scale
(by making some compact volume large \cite{aadd} ) 
in order to avoid a gauge hierarchy \cite{afiru,angel} . 
Alternatively one could consider fully $\cn = 1$ SUSY
extensions of this model, which is anyway $\cn = 1$ 
supersymmetric locally. In particular
it would be interesting 
to see whether one can embed this very simple brane configuration 
into a fully $\cn = 1$ SUSY model as in refs.\cite{csu,bgo}.

\item Irrespective of its construction as a string model,
one may consider the brane configuration itself, with the
fermions and Higgs fields at intersections,
as a phenomenologically viable model for the
understanding of quark and lepton masses and
CP violation. The idea would be  that analogously
simple D-brane configurations incorporating 
similar mechanisms and symmetries could 
possibly be present in large classes of 
models with D6-branes wrapping cycles
on compact manifolds (like, e.g., a Calabi-Yau).
 Note in this respect that the
size of the string scale, either high or low, has
no relevance for the computation of the Yukawa couplings.
Thus our Yukawa coupling formulae may be applicable to
both low string models and more conventional 
models with the string scale of order the GUT scale.
We have seen, however, that in this toroidal example 
the size of the second and third tori have to be
of order of the string scale in order to adjust the 
fermion spectrum data.

\item The  explicit formulae for Yukawa couplings 
provide a new laboratory for the understanding 
of the structure of fermion masses. Some properties 
of these formulae include possibilities phenomenologically 
explored previously. Thus for example, 
our Yukawa couplings depend on the brane location parameters,
the $\epsilon$'s. Those correspond to vacuum expectation values of 
adjoints under the full gauge group (including $U(1)$'s).
Expanding our formulae on the $\epsilon $-parameters 
would give rise to non-renormalizable couplings 
of the form $H\psi_R\psi_L (adjoint)^n$. 
This kind of couplings have been used since a long time
in attempts to understand the SM Yukawa couplings.
Unification symmetries like GUT's have also played an important role 
in previous analysis.
In the  specific example studied in detail a 
Pati-Salam symmetry $SU(4)\times SU(2)_L\times SU(2)_R$
is underlying the brane configuration.
The present  scheme has also some analogies with the
structure of Yukawa couplings in heterotic 
$\int_N$ orbifold models \cite{hv,orbihier}, in which fermions and Higgs
multiplets
located at different fixed points interact via world-sheet
instanton effects\footnote{ From the string theory point of view 
there are a number of differences though. 
For example, in the present case 
the Yukawa couplings come from open strings in the disk.
In the heterotic string there are only closed strings 
and the relevant amplitude appears on the sphere.}.
This is also somewhat analogous to recent attempts
\cite{gauss}  to understand the 
fermion spectrum in terms of extra dimension models in which the 
fermions have different locations (with, e.g., a Gaussian wave 
function) and the different wave function overlap with the Higgs 
field determines the structure of Yukawa couplings.

\item 
Other property that we find is  the presence of 
accidental (approximate) global (e.g., $SU(2)$) symmetries 
in the structure of Yukawa couplings
\footnote{The presence of flavour $SU(2)$ global symmetries 
in the mass matrices have been previously considered, e.g., in
\cite{barbi}.}. 
For some geometrically symmetric brane locations
some of the triangles formed by the branes become identical. 
At those locations (and for real phases) two columns and/or 
rows of the fermion mass matrices become identical, 
signalling the presence of a massless fermion. The numerical
analysis shows that in order to have sufficiently light
up- and down-quarks the branes should be close to these
symmetry points. We have also found that for other symmetric
brane configurations there are cancellations between different
instanton contributions giving rise to
vanishing columns and/or rows 
in the mass matrices. It is the interplay of these 
two effects which turns out to be able to reproduce the
observed fermion spectrum, rather than any exponential suppression
of the lightest fermions.

\item One of the nice insights of the present construction is that of
the origin of CP-violation. In our constructions complex phases
appear associated to the generic presence of non-vanishing
Wilson lines of the $U(1)$'s  of the model. We believe this is 
rather general and is not specific to  our particular constructions. 
This means that the origin of CKM CP-violation is an
open string effect, i.e., it is associated to the D-branes, 
rather than to the bulk fields. This is interesting because
if promoted to a fully $\cn = 1$ setting, 
it may provide a solution to the supersymmetry CP-problem,
that is, why the phases of soft terms are so small (which is
required from EDMN limits) compared to the CKM phase
(for some recent work on CP-violation in string models see, e.g.,
\cite{stringCP} and for the same issue in extra dimensions see
\cite{cojoCP}).
If soft terms physics  
comes from the bulk (closed string) gravitational 
sector, they may be all real without affecting CKM mixing.
In other words CKM CP-violation would come from the open string sector
and soft-term CP-violation from the closed string (bulk) sector.

\end{itemize}

In summary, we have shown how simple configurations 
of D-branes wrapping a compact space
 may give a good quantitative description
of quark masses and mixing angles as well as CP-violation and 
charged lepton masses.
We believe that the approach is more general than the
particular example numerically explored since all realistic D-brane 
models constructed up to now may be re-expressed as coming from
D6-branes wrapping cycles in some compact space \cite{angel}.
 
One of the nice features of this approach is
that one can obtain simple explicit formulae for each Yukawa 
as a sum over world-sheet instanton contributions. 
A non-vanishing value for $U(1)$ Wilson lines lies at the
origin of complex phases and CP violation.
Symmetric 
configurations of the branes lead to approximate symmetries 
of the fermion mass matrices, giving rise to particular
patterns in the fermion spectrum. Interestingly enough 
our analysis shows that the experimental data is
consistent with the existence of an underlying 
Pati-Salam symmetry   which is broken by
shifting of the brane locations (adjoint Higgsing
in the field theory language). This shifting is also the
cause of the reversion of the up-down hierarchy (i.e., 
$m_{up}<m_{down}$), the presence of CKM mixing 
and CP violation. 
It is certainly worthwhile to study the neutrino 
mass spectrum from this perspective. We hope to report
on this issue in the near future.

\vspace{3cm}

\centerline{\bf Acknowledgements}
We are grateful to G. Aldaz\'abal, G. Honecker, C. Kokorelis,
 F. Quevedo, R. Rabad\'an and  A. Uranga for useful discussions.
The research of D.C. and F.M. was  supported by
 the Ministerio de Educaci\'on, Cultura y Deporte (Spain) through FPU grants.
This work is partially supported by CICYT (Spain) and the
European Commission (RTN contract HPRN-CT-2000-00148).

\end{document}